\documentclass[a4paper,fleqn, 11pt]{cas-sc}

\usepackage{setspace}
\usepackage{caption}
\usepackage{subcaption}
\usepackage[numbers]{natbib}
\usepackage{listings}
\usepackage{xcolor}
\usepackage{longtable}
\usepackage[norndcorners,customcolors]{hf-tikz}
\hfsetbordercolor{yellow}
\hfsetfillcolor{yellow!10}

\usepackage{xurl}

\usepackage{todonotes}
\usepackage{seqsplit}

\newcommand{\longstring}[1]{{\ttfamily\seqsplit{#1}}}

\newcommand{\schema}[0]{ARMAND}
\newcommand{\toolname}[0]{ARMANDroid}

\def\tsc#1{\csdef{#1}{\textsc{\lowercase{#1}}\xspace}}
\tsc{WGM}
\tsc{QE}
\tsc{EP}
\tsc{PMS}
\tsc{BEC}
\tsc{DE}

\newboolean{showcomments}
\setboolean{showcomments}{true}         

\ifthenelse{\boolean{showcomments}}
  {\newcommand{\nb}[2]{
  \fbox{\bfseries\sffamily\scriptsize#1}
     {\sf\small$\blacktriangleright$\textit{\textcolor{red}{#2}}$\blacktriangleleft$}
   }
  }
  {\newcommand{\nb}[2]{}
   
  }

\begin{document}
\let\WriteBookmarks\relax
\def\floatpagepagefraction{1}
\def\textpagefraction{.001}
\shorttitle{\schema{}:Anti-Repackaging through Multi-pattern Anti-tampering based on Native Detection}
\shortauthors{A. Merlo, A. Ruggia, L. Sciolla, L. Verderame}

\title [mode = title]{\schema{}: Anti-Repackaging through Multi-pattern Anti-tampering based on Native Detection}


\author[1]{Alessio Merlo}[orcid=0000-0002-2272-2376]
\cormark[1]
\ead{alessio@dibris.unige.it}

\address[1]{DIBRIS - University of Genoa, Via Dodecaneso, 35, I-16146, Genoa, Italy.}

\author[1]{Antonio Ruggia}\ead{antonio.ruggia@dibris.unige.it}

\author[1]{Luigi Sciolla}\ead{luigi.sciolla@dibris.unige.it}


\author[1]
{Luca Verderame}[orcid=0000-0001-7155-7429]
\ead{luca.verderame@dibris.unige.it}

\cortext[cor1]{Corresponding author}


\maketitle
\doublespacing
\begin{abstract}
\emph{App repackaging} refers to the practice of customizing an existing mobile app and redistributing it in the wild to fool the final user into installing the repackaged app instead of the original one. In this way, an attacker can embed malicious payload into a legitimate app for different aims, such as access to premium features, redirect revenue, or access to user's private data. In the Android ecosystem, apps are available on public stores, and the only requirement for an app to execute properly is to be digitally signed. Due to this, the repackaging threat is widely spread.

Anti-repackaging techniques aim to make harder the repackaging process for an attack adding logical controls - called detection node - in the app at compile-time. Such controls check the app integrity at runtime to detect tampering. If tampering is recognized, the detection nodes lead the repackaged app to fail (e.g., throwing an exception). From an attacker's standpoint, she must detect and bypass all controls to repackage safely. 

In this work, we propose a novel anti-repackaging scheme - called \textbf{\schema{}} - which aims to overcome the limitations of the current protection schemes. We have implemented this scheme into a prototype - named \textbf{\toolname{}} - which leverages multiple protection patterns and relies on native code. The evaluation phase of \toolname{} on 30.000 real-world Android apps showed that the scheme is robust against the common attack vectors and efficient in terms of time and space overhead.

\end{abstract}

\section{Introduction}

In the mobile world \textsl{repackaging} refers to the practice of retrieving an app from a public store, decompile it, add some extra code, and then build a new version of the app to deliver in the wild. In the Android ecosystem, repackaging is the main carrier for malware to spread, as an attacker may retrieve a target official app (i.e., Facebook), reverse engineer it \citep{2011-android-reversing}, and enclose a malicious payload. Finally, the attacker packs and signs the modified version of the app (i.e., the repackaged app), and re-distributes it in official app stores (e.g., Google Play Store \cite{google-play-store} and Samsung App Store \cite{samsung-store}) or over the Internet, trying to fool the user to install the repackaged version of the app instead of the original one. This last step can be achieved by well-established attack vectors, like phishing \cite{phishing-ccs18}.  

Albeit in recent years several proposal have been put forward to counteract this threat \cite{7299906, 7579771, 8368344, 8186215, 10.1145/3168820, 10.1007/978-3-030-45371-8_12}, this issue is far from being solved \cite{merlo2020you}, mostly due to the fact that Android carries out just some soft checks on the signature of an app. More precisely, the only requirement for an app bundle (i.e., the $apk$ file) to be installed on a mobile device is that it contains a valid signature, independently from the actual identity of the developer. 
Furthermore, the user has the possibility to install third-party apps from unknown sources, thus limiting the efficacy of security controls carried out on the official app stores (e.g., Google Bouncer \cite{google-bouncer} and Google Play Protect \cite{google-play-protect} for Google Play Store).
Previous choices foster the spread of repackaged apps and demand for proper solutions to counteract this threat.

In this respect, the current SotA is divided into two orthogonal categories, namely i) \textsl{repackaging detection}, that focuses on recognizing and identifying already repackaged apps in the wild, and ii) \textsl{anti-repackaging} (also known as \textsl{repackaging  avoidance}) that aims at protecting original apps against repackaging. 
Briefly, \textsl{anti-repackaging} methodologies focus on inserting logic controls (called \textsl{detection nodes}) that embed integrity checks on the content of the $apk$, called \textsl{anti-tampering} (AT) controls. 
The detection nodes are triggered during the execution of the app, and if some tampering is detected, the app is usually forced to crash. The rationale is to discourage the attacker from repackaging if the likelihood of building a working repackaged app is low.

Concerning anti-repackaging, the scientific community proposed several protection schemes that leverage some fruitful combination of AT controls and detection nodes to protect apps.
Unfortunately, as we demonstrated in a recent work \cite{merlo2020you}, all the proposals so far rely on rigid patterns for inserting detection nodes or adopt only some trivial AT controls that could be easily detected and circumvented by attackers.

To tackle this problem, in this paper we propose a novel anti-repackaging scheme - called \textbf{\schema{}} (\textsl{\underline{A}nti-\underline{R}epackaging through \underline{M}ulti-pattern \underline{A}nti-tampering based on \underline{N}ative \underline{D}etection}) -  that is able to overcome the limitations of the current proposals. \schema{} fulfills the anti-repackaging challenges identified in \cite{merlo2020you} as it relies on multiple protection patterns, heterogeneous AT controls, multi-level detection nodes, and self-hiding techniques. 
\schema{} allows developers to include novel, hard-to-identify and hard-to-reverse security controls in their apps to prevent repackaging.

Furthermore, we developed a prototype implementation for Android apps - \textbf{\toolname{}} - that can be included in the standard development life-cycle of an $apk$. The tool has been evaluated on 30000 Android apps downloaded from different app stores, such as F-Droid \cite{f-droid} and Google Play Store \cite{google-play-store}. The evaluation phase proved the reliability of \schema{} that allows distributing a high number of detection nodes in the apps at the cost of a negligible runtime overheads in terms of memory and cpu usage. To prove the latter result, we tested 220 apps to compare the runtime overhead between original and protected apps. The \toolname{} tool is publicly available on GitHub at \url{https://github.com/Mobile-IoT-Security-Lab/ARMANDroid}, while a Docker image can be found at \url{https://hub.docker.com/r/totor13/armand}.

\paragraph{\textbf{Structure of the paper.}} First, we introduce the threat model related to the app repackaging in the Android ecosystem and the main app-repackaging techniques (Section \ref{sec:background}). 
In Section \ref{sec:related_works}, we report state-of-the-art anti-repackaging protection schemes, thereby underlying their current limitations. 
Then, we present the \schema{} protection scheme, its distinguish features (Section \ref{sec:methodology}), and its resiliency against the main repackaging attacks (Section \ref{sec:security}). 
In Section \ref{sec:implementation}, we present the \toolname{} implementation and, in Section \ref{sec:experimental_result}, we detail the experimental campaign on 30.0000 real-world Android apps and the run-time evaluation of 220 protected apps.
Finally, in Section \ref{sec:conclusion} we conclude the paper by summing-up the main takeaways and putting forward some consideration for future works.

\section{Background \& Threat Model} \label{sec:background}

This section summarizes the basics of app development and distribution in mobile world with a specific focus on Android, a threat model for app repackaging, and the main techniques at the basis of anti-repackaging schemes.

\subsection{Android App  Life-cycle}
An Android app is usually developed in Java or Kotlin, and leverages XML files to represent the graphical user interface of the app. In addition, some developers can implement parts of their app in Native code (i.e., C/C++ code) leveraging the Android NDK toolset \cite{android-ndk}. 
Each app is distributed and installed as an Android Package ($apk$) file, which is a zip archive containing the executable code in a set of \texttt{classes.dex} files, some binary resources (e.g., picture and movies) as well as a \textsl{manifest} file containing, among others, the description of the app components and the permissions required by the app to execute properly. Finally, the $apk$ also includes some meta-files containing the signature of the app, the public-key certificate of the developers and some file for redundant integrity check. 

Android apps are mainly delivered through software repositories, called \textsl{app stores}, which allow the developers to publish their $apks$, and the users to search, download and install them on their devices.
In addition, it is also possible to distribute an $apk$ directly (e.g., through email or URLs) without the need to upload the app to an app store.

Android requires all $apks$ to be digitally signed in order to be installed on a device.
This mechanism grants the integrity of the $apk$: during the installation phase, Android checks the signature of the $apk$ according to the public-key certificate therein, and denies the installation in case either the signature or the integrity check fail.
However, no identification of the signer is required, thus allowing the installation of apps with self-signed or unknown certificates \cite{android-publish-app}.

\subsection{Threat Model}\label{sub:threat-model}
\begin{figure}[!ht]
  \includegraphics[width=\textwidth]{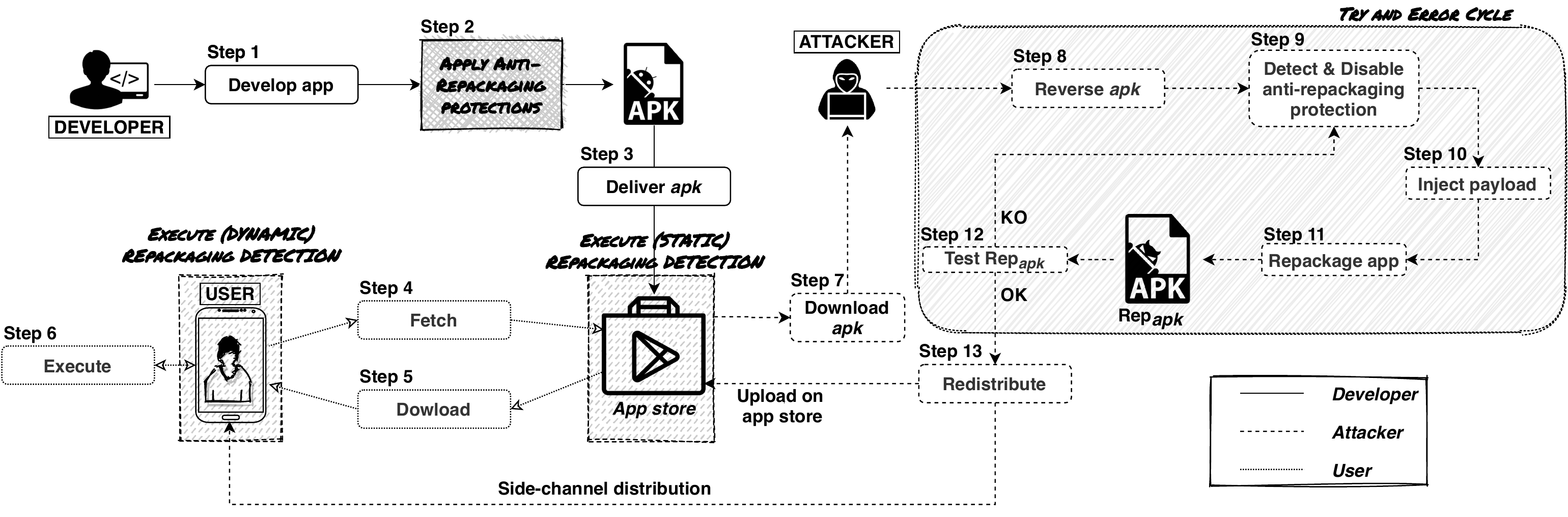}
  \caption{App repackaging: threat model.}
  \label{fig:threat-model-design}
\end{figure}

The threat model for app repackaging involves three actors: the developer, the attacker, and the user.
In an ideal (i.e., attacker-free) scenario, the developer implements the app (Step 1), signs it, and delivers the app to the users via one or more app stores (Step 3). 
The user fetches the app (Step 4) and downloads it (Step 5). Android verifies the signature and the integrity of the $apk$ and installs it on the user's device. Then, the app can execute (Step 6). 

Nonetheless, in a real scenario the developer applies some anti-repackaging protection schemes during the app development (Step 2), to deter the attacker to repackage the app. From her side, the attacker accesses the app directly from the app store, behaving like any regular user: she downloads the app (Step 7), decompiles and reverses the $apk$ (Step 8), and tries to detect and disable potentials anti-repackaging (Step 9). Then, the attacker add some malicious payload to the app (Step 10) and builds-and-signs a repackaged version (i.e., $Rep_{apk}$) of the original app (Step 11). Finally, the attacker tests whether $Rep_{apk}$ works properly (Step 12): if this is the case, the attacker redistributes $Rep_{apk}$ both on app stores and through side-channels (e.g., by email, web servers, or leveraging phishing attacks), aiming to force the user to install $Rep_{apk}$ instead of the original $apk$.  
Otherwise, the attacker tries further to detect and disable other anti-repackaging protections on the original $apk$ (back to Step 8). Steps 8 to 12 are also known as the \textsl{try and error cycle} that the attacker must keep executing until she gets a working repackaged app. To this aim, an ideal anti-repackaging scheme never allows the attacker to obtain a working repackaged app (i.e., moving out from the \textsl{try and error cycle}). Actually, a reliable anti-repackaging solution makes the repackaging non cost-effective, i.e., it requires so much time to be disabled that the attacker gives up from repackaging the app. 

As a final observation, \textsl{repackaging detection techniques} are instead applied on the app store and, sometimes, on the mobile device to detect already repackaged apps. However, in this paper we will focus on anti-repackaging only. 

\subsection{Anti-repackaging Techniques}\label{sub:anti-repackaging}

Anti-repackaging - also known as \textsl{repackaging avoidance} or \textsl{self-protection} - aims to protect an app from being successfully repackaged by adding some protection code - called \textsl{detection nodes} - in the app code before building the $apk$ and delivering it to the app store (Step 2 in Fig. \ref{fig:threat-model-design}). 
The idea of detection nodes has been put forward in \cite{7579771} and it refers to a self-protecting mechanism made by a piece of code inserted into the original $apk$, which carries out integrity checks - called \textsl{anti-tampering} controls (e.g., signature check, package name check) - when executed at runtime \cite{merlo2020you}.
More specifically, anti-tampering checks compare the signature of a specific part of the $apk$ with a value pre-computed during the building of the original app; if such values differs, then a repackaging is detected and the detection node usually leads the app to fail, thereby frustrating the repackaging effort.
 
From the attacker side, most of the activities related to detect and disable anti-tampering refers to the identification and removal of detection nodes. 
To protect detection nodes, most of the anti-repackaging schemes hide them into the so-called \textsl{logic bombs} \cite{10.1145/3168820, 8186215, 10.1007/978-3-030-45371-8_12}, that has been originally conceived in the malware world to hide malicious payloads \cite{sharif2008impeding}. A logic bomb protects (i.e., hides) a detection node by ciphering the node using reliable ciphers at the state of the art. An actual value of a program variable is often used as the key. Logic bombs rely on the information asymmetry between the developer and the attacker, i.e., since the attacker has partial knowledge of the app behaviors, it is unlikely that she can correctly guess the key value used to encrypt the code. 
\begin{figure}[!ht]
    \centering
    \includegraphics[width=\textwidth]{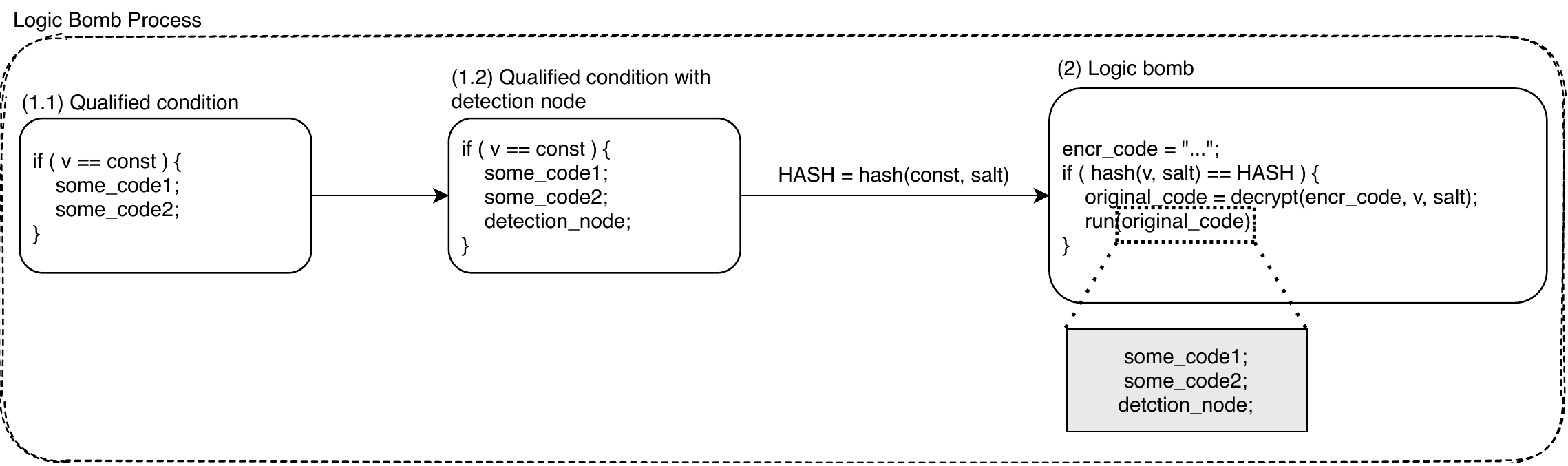} 
    \caption{Transformation process of a qualified condition into a logic bomb.}
    \label{fig:qc_to_lb}
\end{figure}

Figure \ref{fig:qc_to_lb} shows a common scheme for building a logic bomb by re-writing a \textsl{qualified condition} - i.e., a condition containing an equality check, where one of the operands is a constant value (e.g., \longstring{v==const}) - in a branch. 
The original condition is transformed in a new one where the pre-computed hash value of the constant \texttt{const} (i.e., \texttt{HASH}) is compared with the result of the hash function applied to variable \texttt{v} plus some \texttt{salt}. 
The detection nodes are embedded into the body of the qualified condition, which is encrypted using the original \texttt{const} as the encryption key. The protection of the logic bomb is granted by the one-way property of cryptographic hash functions, that makes hard for the attacker to retrieve the original \texttt{const} value (i.e., the decryption key) from its hash. In addition, the anti-tampering checks can be embedded either in Java or in native code (i.e., C/C++) to improve reliability.

From an attacker standpoint, all logic bombs must be detected and removed to create a fully working repackaged $apk$. To this aim, an attacker can leverage both static and dynamic analysis to locate and deactivate the anti-tampering controls (e.g., code analysis and pattern matching, code instrumentation). As a complete analysis of such attacking techniques is out of scope in this paper, the interested reader may refer to \cite{merlo2020you, 10.1145/2996358, LI201767, GAJRANI202073}.

\section{Related Works} \label{sec:related_works}

In recent years several anti-repackaging schemes have been proposed to protect Android apps from repackaging, relying on external AT checks or internal AT agents.

The proposals based on external AT agents include all the techniques that leverage a third-party entity to carry out anti-tampering checks. Examples are remote servers \cite{10.1145/2995306.2995315}, some third-party $apk$ or OS kernel modules \cite{10.1007/978-3-030-05234-8_16}. Albeit promising, those schemes suffer from communication-based vulnerabilities \citep{7299933, 8068748} that could interfere with the protection mechanisms. Furthermore, app developers cannot assume the presence of third-party solutions installed on the device, thus limiting the efficacy of such anti-repackaging proposals in the wild.

Regarding the anti-repackaging solutions based exclusively on internal AT checks, the scientific community proposed six different protection schemes, as surveyed in \cite{merlo2020you}.
The first proposal appeared in 2015 by Protsenko et al. \cite{7299906}. The authors put forward an anti-repackaging scheme that encrypts the \texttt{classes.dex} files in the $apk$, and dynamically decrypts and executes them at runtime. The adoption of a XOR-based encryption of the bytecode is aimed at implementing a tamper-proof checksum. 
In 2016, Luo et al. \cite{7579771} proposed the Stochastic Stealthy Network (SSN). The SSN scheme distributes (and hides) a set of guards into the app source code. When executed, such guards trigger some decision points that use stochastic functions to detect tampering.
Song et al. (2017) \cite{8368344} proposed an app reinforcement framework, named AppIS. The idea of AppIS is to add security units as guards with an interlocking relationship between each other to build redundant and reliable anti-repackaging checks. The guards can be included both in Java and native code.
Chen et al. (2018) \cite{8186215} proposed a self-defending code (SDC) scheme, leveraging the information asymmetry between the developer and the attacker on the app code. The idea is to encrypt pieces of source code and decrypt (and properly execute) them at runtime only in case the app has not been repackaged. Unlike the previous works, each piece of code is encrypted with a different key.
Zeng et al. (2018) \cite{10.1145/3168820} introduced the concept of \textsl{logic bombs} as anti-repackaging protections for Android apps. The approach consists of adding cryptographically obfuscated pieces of code that are executed once proper triggers (i.e., logical conditions) are activated. The authors implemented the approach in a tool called BombDroid.
Tanner et al. (2019) \cite{10.1007/978-3-030-45371-8_12} proposed an evolution of previous schemes. Like BombDroid, this work is based on cryptographically obfuscated logic bombs, inserted in candidate methods with a qualified condition. However, the main difference is that integrity checks - and parts of the original app code - are executed in native code.

Unfortunately, we discovered \cite{merlo2020you} that all the prevision proposals can be potentially circumvented by leveraging static and dynamic detection and neutralization techniques such as code analysis and pattern matching, dynamic testing, and symbolic execution. Moreover, we conducted a full-fledged attack on the latest - and more advanced - anti-repackaging scheme whose code is available on GitHub. 
\section{The \schema{} Scheme} \label{sec:methodology}

In this work, we present \schema{} (\underline{A}nti-\underline{R}epackaging through \underline{M}ulti-pattern \underline{A}nti-tampering based on \underline{N}ative \underline{D}etection), a novel anti-tampering protection scheme that embeds logic bombs and AT detection nodes directly in the $apk$ file.
\schema{} aims at overcoming the main limitations of the current state-of-the-art schemes briefly introduced in Section \ref{sec:related_works}, by addressing the following challenges:

\begin{enumerate}
    \item \textbf{Use of multiple patterns} to disseminate and hide logic bombs in the app. Current techniques adhere to a single common pattern (i.e., qualified conditions and adoption of hash functions as shown in Figure \ref{fig:qc_to_lb}), that can be detected and de-activated by an attacker through static code analysis and pattern matching techniques. To this aim, \schema{} includes several - distinct - patterns for logic bombs and AT tampering checks to harden the detection and removal phase.
    \item \textbf{Rely on native technologies (C/C++)} to hide the anti-tampering protection. As reversing and instrumenting native code significantly increase the detection and removal of security controls, \schema{} introduces a set of logic bombs and AT checks specifically designed to exploit native code.
    \item \textbf{Optimize the effectiveness of the detection nodes}. Some approaches implement few controls in the name of performance, while others embed many bombs that are not actually triggered at runtime. In this respect, \schema{} adopts an effective deployment of logic bombs, ensuring that logic bombs and AT checks are not included in unused code.
    \item \textbf{Use Honeypot bombs}. The idea is to include fake logic bombs that - disguised as legitimate ones - increase the complexity of the detection and removal phases. Although such a concept has already been proposed in the literature, \schema{} is the first methodology that relies on their usage.
    \item \textbf{Hide the anti-repackaging protection}. It is fundamental to improve the stealthiness of anti-repackaging controls and detection nodes to make  hard for the attacker to understand that the reverse-engineered app includes a protection scheme. Therefore, \schema{} relies on hiding techniques, e.g., obfuscation \cite{aonzo2020obfuscapk}, to mask the detection nodes.
\end{enumerate}

In the remaining of this section, we detail the main features of \schema{} and presents the deployment of the protection scheme and the hiding strategies. Finally, we describe the runtime behavior of our anti-repackaging scheme.

\subsection{Multi-patterning Logic Bombs \& AT checks}
\label{subsec:spatternization} 

As logic bombs are currently built according to the canonical pattern depicted in Fig. \ref{fig:qc_to_lb}, the attacker can systematically detect any potential location of the anti-tampering detection nodes by identifying all branches containing hash values in their condition, and then automate their removal. 
To mitigate such an issue, \schema{} adopts six different types of logic bombs:
\begin{enumerate}
    \item \textbf{Java bomb}. It represents the canonical logic bomb implemented in Java code. The inner code is enriched with one or more Java AT checks and then encrypted with the value of the qualifying condition (\texttt{const}). 
    \item \textbf{Native-key bomb.} It is built by encrypting the inner code of the qualified condition where the decryption key is the return value of an anti-tampering check that is implemented in the native code. 
    Figure \ref{fig:native_encryption_procedure} shows the creation process of this type of bomb. 
    
     \begin{figure}[!ht]
      \includegraphics[width=1.05\textwidth]{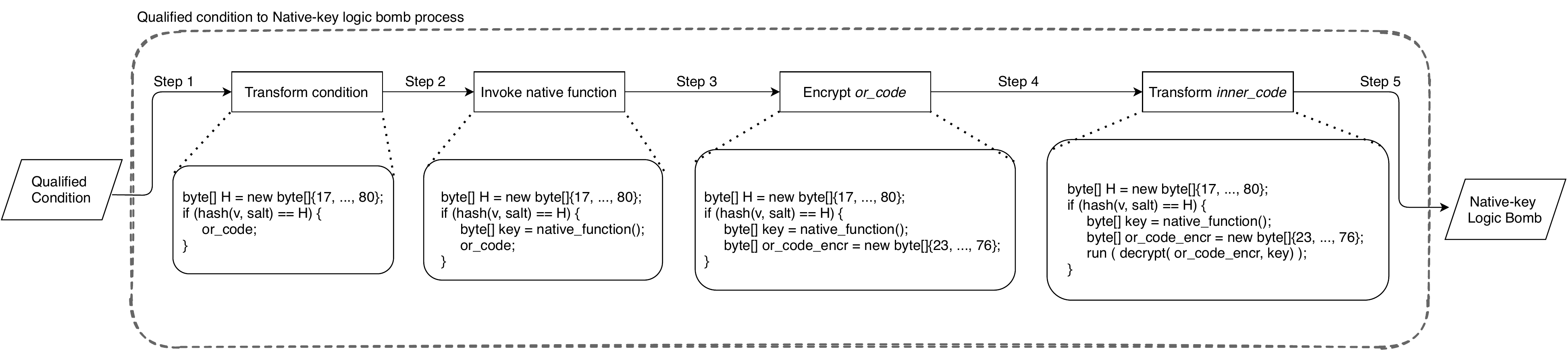}
      \caption{Creation process of a Native-key bomb.}
      \label{fig:native_encryption_procedure}
    \end{figure}

    First, the qualifying condition is transformed into a hash comparison using the $const$ value and a salt (Step 1) generated by a pseudo-random function at compile time. The inner code of the qualifying condition is substituted with a native call (Step 2) whose return value is used to encrypt the original inner code (i.e., \texttt{or\_code}) (Step 3). Then, the bomb includes an instruction to decrypt and run \texttt{or\_code\_encr} (Step 4).
    Finally, the entire \texttt{inner\_code} is encrypted with the \texttt{const} value of the qualified condition to generate the \texttt{encr\_code} (Step 5).
    \item \textbf{Native AT bomb.} It implements one or more anti-tampering checks inside the native code. In detail, \schema{} adds an invocation to a native function in the \texttt{or\_code} that implements one or more native AT checks. Finally, the entire \texttt{inner\_code} is encrypted with the \texttt{const} value of the qualified condition to generate the \texttt{encr\_code}. 
    \item \textbf{Java AT \& Native-key bomb.} This bomb extends the Native-key bomb with AT checks in Java code. To do that, \schema{} includes additional instructions in the \texttt{or\_code} to execute one or more AT checks. The code is then encrypted using the workflow of the native-key bomb.
    \item \textbf{Java \& Native AT bomb.} This category of bombs combines AT checks executed in Java and in native code. The bomb enriches the \texttt{or\_code} with instructions to execute one or more Java AT checks and with a native function invocation that executes one or more native AT checks. Then, the entire \texttt{inner\_code} is encrypted with the \texttt{const} value of the qualified condition to generate the \texttt{encr\_code}.
    \item \textbf{Honeypot bomb.} This is a fake logic bomb, which does not contain any anti-tampering check (i.e., the \texttt{or\_code} is not changed). Honeypots aim at tricking the attacker into spending time and effort to decrypt them uselessly.
\end{enumerate}

In our work, each anti-tampering node executes a different AT technique to detect a \emph{specific type} of repackaging. 
\schema{} supports several controls in either Java and native code, such as environment controls (e.g., emulator detection), signature check, package name check, and file integrity check (e.g., resource and image files). We refer the interested reader to \cite{merlo2020you} for a detailed description of the different AT techniques that can be used in an anti-repackaging scheme.

In addition to logic bombs, \schema{} also allows the inclusion of additional detection nodes, called \textbf{Java AT}. Such anti-tampering checks are spread in the entire Java code of the app to detect potential repackaging attempts and are not included in logic bombs. To this aim, the attacker needs to scan and inspect the entire code (and not only the qualifying conditions) to detect and bypass each of them.

\subsection{Protection Deployment and Hiding Strategies}

State of the art anti-tampering schemes tend to add multiple detection nodes that share a single implementation of an AT check (e.g., a signing verification) located either in the Java or native part of the app. Consequently, disabling a single AT check (e.g., using dynamic code instrumentation) could imply the removal of several (or all, in the worst case) detection nodes, thus invalidating the protection.

To mitigate the problem, \schema{} replicates - for each logic bomb or AT check - the corresponding code, and encrypts each piece of code with a different key of the bomb. Such a deployment strategy allows the mitigation of single-point-of-failure exploits since the attacker needs to detect and disable every single control included in the $apk$ file.

Furthermore, \schema{} extends the concept of a logic bomb to \textbf{nested logic bomb}, i.e., where the content of a logic bomb can hide one or more encrypted bombs. In this way, an attacker has to decrypt all the nested logic bombs to recover and remove the included anti-tampering checks.
Besides, \schema{} inserts a different type of bomb for each qualified condition and a different AT controls each time, based on semi-stochastic criteria. Briefly, the \schema{} anti-tampering scheme allows the definition of a set of probability values (one for each type of bomb), which are used to determine the type of logic bombs that will be embedded. Also, \schema{} includes a percentage indicating the frequency with which the Java AT node is spread through the bytecode to increase the complexity of the protection profile. This process is detailed in Section \ref{sec:implementation}.

Finally, \schema{} empowers a set of obfuscation engines, based on the Obfuscapk methodology \cite{aonzo2020obfuscapk}, to further prevent the detection of logic bombs and AT checks. 

\subsection{Runtime  Behavior}

At runtime, the Java AT checks and the logic bombs included in the app are executed. The code in the logic bombs is executed if and only if the hash value of the variable (\texttt{v}) in the qualifying condition (combined with the pseudo-random salt) is equal to the hard-coded hash (i.e., the pre-computed hash value of \texttt{const}). 
If this is the case, the inner code is decrypted using \texttt{v} as the decryption key.
Such a code could contain: i) the original instructions and no anti-tampering checks (Honeypot bombs), ii) one or more Java AT checks - that are immediately executed, iii) a call to a native function (e.g., Native anti-tampering controls), or iv) an inner logic bomb. In the latter case, the execution will proceed iteratively to decrypt the inner bomb and act accordingly. When an AT check is triggered, the code immediately runs the anti-tampering control (e.g., a signature verification). In case of failure, the AT check raises an exception that blocks the execution of the app.

\begin{figure}[!ht]
    \centering
    \includegraphics[width=1.02\textwidth]{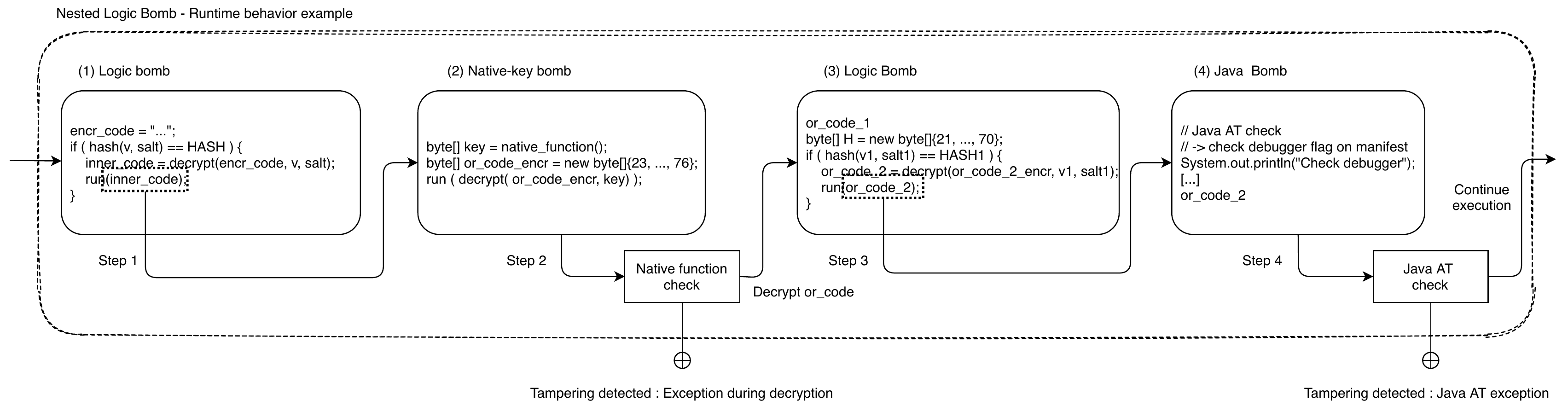}
    \caption{Runtime behavior of a nested logic bomb.}
    \label{fig:LB-runtime-behavior}
\end{figure}

Figure \ref{fig:LB-runtime-behavior} depicts an example of two nested logic bombs: a \emph{Native-key bomb} that contains a \emph{Java bomb}.
Initially, the app evaluates a first logic (canonical) condition ($hash(v, salt) == HASH$). If the equality is satisfied, 
the inner code (i.e., \texttt{encr\_code}) of the logic bomb is decrypted and executed (Step 1). 
The \texttt{inner\_code} contains i) a call to a native function that returns the correct decryption key if a tampering control succeeds and ii) a ciphered code (i.e., \texttt{or\_code\_encr}). 
In this way, if a repackaging is detected, the decryption function fails (i.e., the decryption key is not correct), and the app crashes, throwing an exception; otherwise, the decrypted code is executed (Step 2). 
Part of the decrypted code (i.e., \texttt{or\_code\_1}) can be executed directly, while the remaining part corresponds to a Java bomb.
Thus, the app evaluates the second logic condition ($hash(v1, salt1) == HASH1$) (Step 3). In case of success, the code is decrypted to obtain \texttt{or\_code\_2}. Finally, the app executes the Java AT checks included in the Java bomb (in the example, it checks the debugger flag). If such a control detects a repackaging, it will throw an exception (causing an app crash); otherwise, the app continues the execution of \texttt{or\_code\_2} (Step 4).

\section{Security Analysis}\label{sec:security}

According to the threat model in Section  \ref{sub:threat-model}, the goal of anti-repackaging is to include robust security controls that the effort required to remove them (Step 9) may be great enough to discourage the attacker from repackaging.

In such a scenario, the attacker has only two strategies to repack an app successfully. In the first case, she needs to detect and remove all the detection nodes spread in the app code. The second possibility is to tamper with the AT checks to avoid the repackaging detection (e.g., by overriding the return value of an anti-tampering function). 
Both strategies can leverage a \textsl{weapon rack} of techniques to analyze a protected $apk$ to detect and dismantle anti-repackaging protections. 
The weapon rack includes several static and dynamic techniques such as code analysis and pattern matching, dynamic analysis \& testing, and code manipulation techniques \cite{merlo2020you}.

In the rest of this section, we evaluate the reliability and robustness of \schema{}, explaining the difficulties for an attacker to deactivate the protection.

\subsection{Detection Nodes: Identification and Removal} 

All the six types of logic bomb in \schema{} hide under the same structure ($hash(v, salt) == H$). Nonetheless, each bomb has a different content, i.e., it can optionally i) include one or more AT checks, ii) rely on native code, and iii) contain other nested logic bombs.
Thanks to such a diversity, the attacker cannot infer a common pattern for the logic bombs, making difficult the adoption of automated detection and/or removal process. 
Furthermore, even if the attacker has the ability to detect all the patterns related to the logic bombs, she still has to find and circumvent the AT checks that are randomly spread in the app code and are external to any logic bomb.

The \schema{} protection scheme increases also the resiliency against dynamic analysis and automated testing techniques.
For instance, an attacker can leverage code instrumentation to intercept the call to the decryption function (e.g., AES decrypt) inside a logic bomb to recover the decryption key and the \texttt{inner\_code}.
However, thanks to the multi-pattern strategy adopted by \schema{}, an attacker has to manually analyze the content of each logic bomb to neutralize or remove the nested Java and/or native AT controls. 
Furthermore, the inclusion of nested logic bombs provide a higher level of protection: even if an attacker decrypts the first level, she needs to iterate the process to remove AT checks until she reaches the \texttt{or\_code}.

Under previous conditions, the attacker has to \emph{manually} execute most of the app code to be sure that all detection nodes are removed and bypassed, making the repackaging process extremely time and resource consuming, in particular if the \schema{} implementation distributes a lot of bombs, pervasively.

\subsection{Deactivation of AT checks} 

The attacker may avoid looking for detection nodes, and focus on deactivating the AT checks they invoke to verify the app integrity. \schema{} relies on several types of anti-tampering controls (e.g., signature checking and file integrity) selected using a pseudo-random heuristic, which is adopted both for Java and native code. 
In this way, an attacker has to find a different approach for each type (and implementation) of an AT check. Furthermore, the reverse engineering and instrumentation process of native code (assembly code) requires additional technical skills that increase the difficulty of the attack w.r.t. the Java part.
Also, \schema{} relies on cross-language AT controls, where the computation is split in both Java and native code. Such a dependency avoids the usage of \emph{code deletion} attacks since removing a native call invocation would break the expected execution of the Java code, leading the app to crash.
Finally, the code of the anti-tampering controls is replicated for each usage. In this way, the attacker must detect and tamper with all the AT check invocations to bypass the anti-tampering control.

\section{\toolname{}: an implementation for Android of \schema{}} \label{sec:implementation}

We implemented the \schema{} anti-repacking scheme in a prototype tool called \toolname{}, that can be executed as a Docker image available at \url{https://hub.docker.com/repository/docker/totor13/armand}.
The tool is transparent to the development pipeline as it can work directly on pre-delivery $apk$ files. Indeed, the developer can build the app bundle, apply the \toolname{} protection, and finally sign the resulting app for the distribution.


\toolname{} is written in Java and relies on the Soot framework \cite{soot} to analyze and modify the Java bytecode, 
 and the Axml\footnote{\url{https://github.com/Sable/axml}} library to parse and modify the Android manifest file.
The tool exposes four input parameters, i.e., three percentage values ($P_{Java\_{AT}}$, $P_{native\_key}$, and $P_{native\_AT}$) and a package name (\texttt{PN}).
Briefly, the three percentage values allow \toolname{} to determine the probability to include the Java AT checks and the six types of logic bombs. 
The package name specifies the part of the app in which the tool will inject the protection controls. If \texttt{PN} is set to $none$, the tool will apply the protection to all the Java classes in the app. 

\begin{figure}[!ht]
    \includegraphics[width=\textwidth]{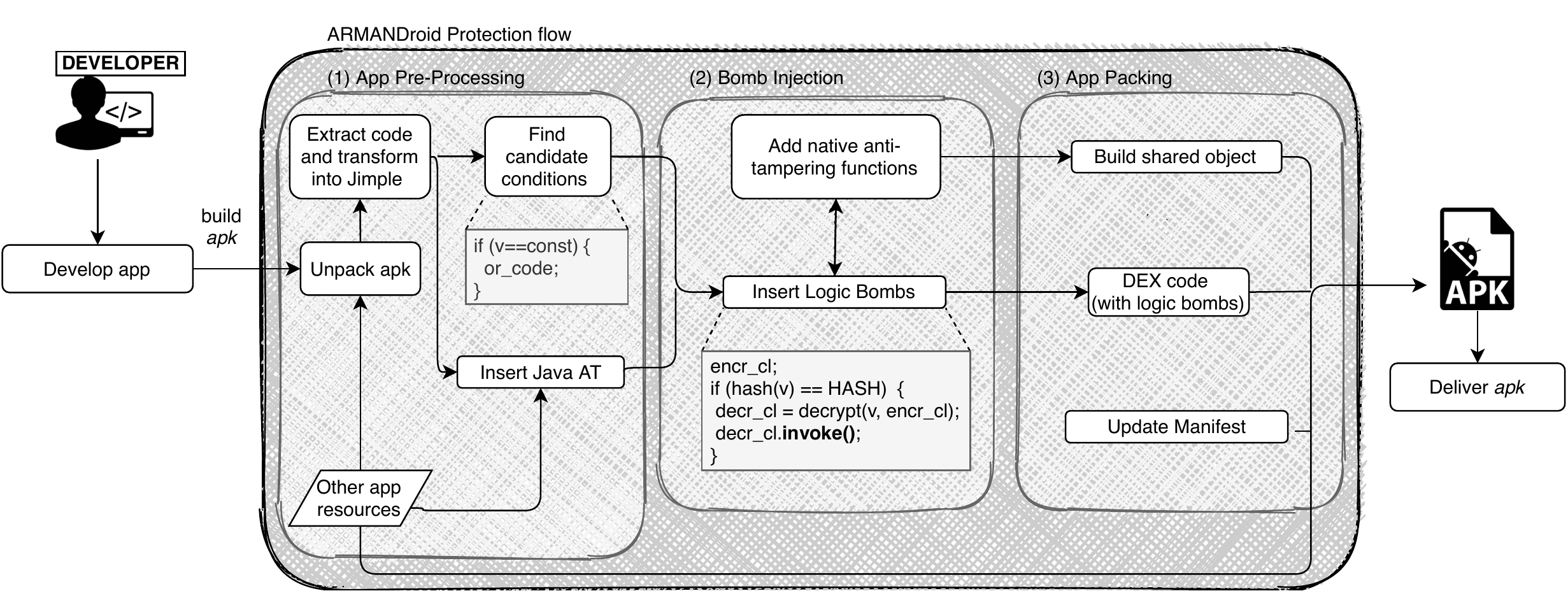}
    \caption{The \toolname{} protection workflow.}
    \label{fig:antirepackaging-protection-flow}
\end{figure}
The protection process can be divided into three macro-steps (see Figure \ref{fig:antirepackaging-protection-flow}):
\begin{enumerate}
    \item \emph{App Pre-processing.} \toolname{} unpacks the $apk$ file to extract the \texttt{classes.dex} file(s). The bytecode is then translated into an intermediate language, called Jimple \cite{vallee1998jimple}, used by Soot for analysis and modification tasks.  During this phase, \toolname{} scans the app code in \texttt{PN} to find all the qualified conditions (QC).
    Furthermore, the tool injects a set of Java AT checks into the bytecode. 
    \item \emph{Bomb Injection.} For each QC, the tool includes - according to the input probabilities $P_{Java\_AT}$, $P_{native\_key}$, and $P_{native\_AT}$ - one or more logic bombs belonging to the six different types defined in Section \ref{subsec:spatternization}. If the bomb requires native code, \toolname{} builds the corresponding native method and all the auxiliary code needed for its execution.
    \item \emph{App Packing.} Finally, \toolname{} transforms back the Jimple representation into bytecode, adds the native libraries, updates the \longstring{AndroidManifest.xml} file, and creates the protected $apk$ file, ready for the signing process.
\end{enumerate}

The rest of this section details the implementation of each of the steps.

\subsection{App Pre-Processing}\label{subsec:app-analysis}

During this phase, the tool processes the app's code to i) identify all the QCs that can host logic bombs and ii) inject the Java AT controls.

\paragraph{\textbf{Unpacking and translation.}} At first, \toolname{} unpacks and extracts from the $apk$ file the \texttt{classes.dex} file(s) containing the compiled Java code. The compiled code is then transformed in the Jimple intermediate language using Soot.

\paragraph{\textbf{QC Identification.}} \toolname{} scans the Jimple code belonging to the package $PN$ to detect sets of consecutive instructions (hereafter, blocks) that can host one or more logic bombs. These blocks must match the \emph{qualified condition} pattern as described in Section \ref{sub:anti-repackaging}.

In the current implementation, \toolname{} considers two types of QCs, i.e., if statements and switch-case statements. For the latter case, each case of the switch instruction is translated into a separate qualified condition. 
For each discovered QC, \toolname{} builds an Abstract Syntax Tree (called \emph{transformation tree}) with the qualified condition as the root node.
In the case of nested QCs, the transformation tree will add the inner QCs as leaf nodes to keep track of their hierarchy.

Moreover, the tool also stores the link to the beginning and the end of the block of instructions that can be encrypted in the bomb, namely the \emph{encryption range}, for each discovered QC. The encryption range encloses the set of consecutive instructions (up to the maximum size of \texttt{or\_code}) that do not contain further branches. 

\paragraph{\textbf{Java AT Injection.}} During the parsing phase, \toolname{} injects Java AT controls into the app code. In detail, the tool injects a Java AT before each parsed instruction with the input probability $P_{Java\_AT}$, using the algorithm sketched in Figure \ref{fig:add-java-at}.
At the end of the process, the probability to embed at least a Java AT into a set of \texttt{n} instructions is $P_{Java\_AT\_embedded} = 1 - (1-P_{Java\_AT})\;^{n}$.

\begin{figure}[!htbp]
    \begin{lstlisting}[language=Java,
    basicstyle=\footnotesize\ttfamily,
    numbers=left,
    stepnumber=1,
    showstringspaces=false,
    tabsize=1,
    breaklines=true,
    breakatwhitespace=false,
    ]
    Iterator<Unit> iter = units.snapshotIterator();
    Unit currentUnit = iter.hasNext() ? iter.next() : null;
    Unit nextUnit = null;
    while (iter.hasNext()) {
        nextUnit = iter.next();
        Stmt stmt = (Stmt) currentUnit;
        
        [...]
        
        if (!(stmt instanceof IdentityStmt) && 
                random.nextInt(100) >= (100-P_JAVA_AT)) {
            // add Java AT before stmt
            [...]
        }
        currentUnit = nextUnit;
    }
    \end{lstlisting}
\caption{Code snippet adding Java anti-tampering checks.}
\label{fig:add-java-at}
\end{figure}

\subsection{Bomb Injection} \label{sub:code_transformation}

\begin{figure}[!ht]
    \includegraphics[width=\textwidth]{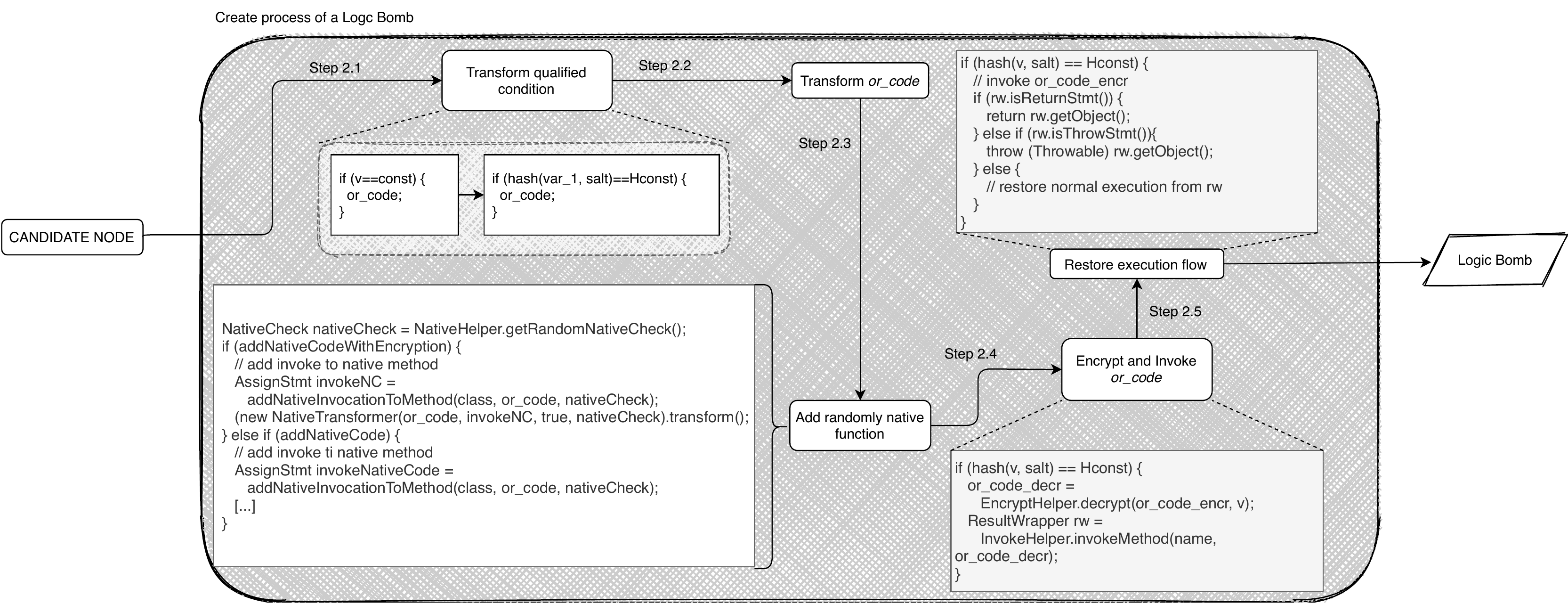}
    \caption{Transformation of a candidate node into a logic bomb.}
    \label{fig:creation-of-logic-bomb}
\end{figure}

In the Bomb Injection phase, \toolname{} traverses each transformation tree using a post-order procedure to process all the QCs detected in the previous phase. The goal of this phase is to transform each QC into a logic bomb belonging to one of the six types defined in Section \ref{sec:methodology}, using the process depicted in Figure \ref{fig:creation-of-logic-bomb}.  

At first, the qualified condition is transformed into 
$hash(v, salt) == H$, where the \texttt{salt} is a pseudo-random value and $H$ is the pre-computed hash value of the constant \texttt{const} (Step 2.1). Step 2.2  evaluates the \texttt{or\_code} to ensure that the original method will be correctly executed after the transformation process. To do so, \toolname{} inspects the execution flow to detect how the execution will behave after the execution of \texttt{or\_code}.
In particular, we identify four possible situations, as the original code may i) proceed to the next block of instructions, ii) throw an exception, iii) jump to a different block, or iv) terminate the execution returning an object.
Thus, the tool - similarly to the solution proposed in  \cite{10.1007/978-3-030-45371-8_12} - adds an auxiliary Java class called \longstring{ResultWrapper} in charge of enforcing the original execution flow after \texttt{or\_code}.

After the first two steps, \toolname{} evaluates the content of \texttt{or\_code} and injects a native call implementing either a Native-key or a Native AT function (Step 2.3).
The probability to create a specific type of a logic bomb depends on the input percentages $P_{native\_key}$ and $P_{native\_AT}$, and the content of the \texttt{or\_code}.

In detail, if the \texttt{or\_code} contains at least one (or more) \texttt{Java AT} (with probability $P_{Java\_AT\_embedded}$), the resulting bomb will be a \texttt{Java AT \& Native-key} bomb with a probability of $P_{native\_key}$, while the probability to have a \texttt{Java \& Native AT bomb} is $P_{native\_AT\; |\; \neg\; native\_key}$. 
If no native functions are injected, then the QC will contain a plain \texttt{Java Bomb} with the probability $P_{Java\_bomb} = 1 - P_{native}$ where $P_{native}$ corresponds to the join probabilities of having a logic bomb with a \emph{native key} or a native AT control (i.e., $P_{native} = P_{native\_key}\; \cup \;P_{native\_AT\; |\; \neg\; native\_key}$).

Otherwise, if \texttt{or\_code} does not contain any Java AT (with probability $1 - P_{Java\_AT\_embedded}$), the bomb could be a \texttt{Native-key bomb} (with probability $P_{native\_key}$), a \texttt{Native AT} bomb (with probability $P_{native\_AT\; |\; \neg\; native\_key}$), or - if no injection occurred - a \texttt{Honeypot}.

In Step 2.4, \toolname{} encrypts \texttt{inner\_code} in the \emph{encryption range} using AES-128 and the result is included in an encoded string using $base64$ (\texttt{inner\_code\_encr}). The encryption/decryption key is derived from the value of \texttt{const}. 
Then, \toolname{} adds the  instructions required to decrypt and invoke the encrypted inner code. To this aim, the tool embeds two auxiliary classes in the Android app, i.e., \longstring{EncryptHelper} and \longstring{InvokeHelper}. The first one is responsible to decrypt \texttt{inner\_code\_encr} using the decryption key. The latter class is responsible for loading a byte array (corresponding to the bytecode resulting from the decryption of the \texttt{inner\_code\_encr}) into the memory and invoking the $execute$ method. In particular, \longstring{InvokeHelper} leverages the \longstring{InMemoryDexClassLoader} class\footnote{\url{https://developer.android.com/reference/dalvik/system/InMemoryDexClassLoader}} to load and execute the code from a byte array (i.e., the byte of the dex file). 
Finally, in Step 2.5, the tool recovers the original execution flow by invoking the \longstring{ResultWrapper} object. 

\subsection{App Packing}

The last step enables the generation of a working protected $apk$ file. 
At the beginning, the Jimple representation of the code is transformed back into the Java bytecode. 
After that, \toolname{} generates a set of shared libraries, one for each supported ABI, containing the native functions included by the tool in the previous phase. The current implementation of the tool supports \texttt{armeabi-v7a}, \texttt{arm64-v8a}, \texttt{x86}, and \texttt{x86-64}.

Then, \toolname{} updates the \longstring{AndroidManifest.xml} file to include the auxiliary classes and the new shared libraries and builds the new $apk$ file.
Finally, \toolname{} sends the protected app to the Obfuscapk \cite{aonzo2020obfuscapk} tool to mask the anti-tampering protections included in the app. 
At the end of this phase, the output \texttt{apk} is ready for the signing process. 
\section{Empirical Evaluation} \label{sec:experimental_result}
 
We empirically assessed the reliability of \toolname{} by applying the tool over a set of 30.000 real-world Android apps downloaded from F-Droid \cite{f-droid} and Google Play \cite{google-play-store} between August and October 2020 (cf. Section \ref{sec:experiments}). Then, we evaluated the best set of input parameters using a subset of 1.000 randomly-selected apps (cf. Section \ref{sec:parameter_tuning}). Finally, we compared the run-time performance between the standard and the protected version on a subset of 220 apps to evaluate the computational overhead and the effectiveness of the \schema{} protection scheme (cf. Section \ref{subsec:effectiveness}).


\subsection{Experimental results}\label{sec:experiments}

The experiments were conducted on a Virtual Machine running Ubuntu 20.04 with 16 dual-core processors and 32GB of RAM. The experimental evaluation used the following parameters: $P_{Java\_AT}=10\%$, $P_{native\_encryption}= 40\%$, $P_{native\_AT}=60\%$, and $PN=ApplicationId$. During our test, we enforced a timeout of 1.200 seconds for the protection process.

\toolname{} was able to successfully protect 27.656 out of 30.000 apps (i.e., 92.2\%) in nearly 14 days and 5 hours (1.228.009 seconds). 
The analysis of the 2.344 remaining apps (i.e., 7.8\%) failed due to the following reasons: 

\begin{itemize}
    \item 986 apps (3.28\%) failed due to well-known issues of the Soot library\footnote{\url{https://github.com/soot-oss/soot/issues/1413}, \url{https://github.com/soot-oss/soot/issues/1474}};
    \item 963 apps (3.21\%) crashed during the tool transformation process;
    \item 229 apps (0.76\%) reached the timeout for the processing;
    \item 166 apps (0.55\%) crashed due to an issue of the Axml library \footnote{\url{https://github.com/soot-oss/soot/issues/1070}}.
\end{itemize}

\begin{figure}
\centering
\begin{subfigure}{.5\textwidth}
  \centering
  \includegraphics[width=\linewidth]{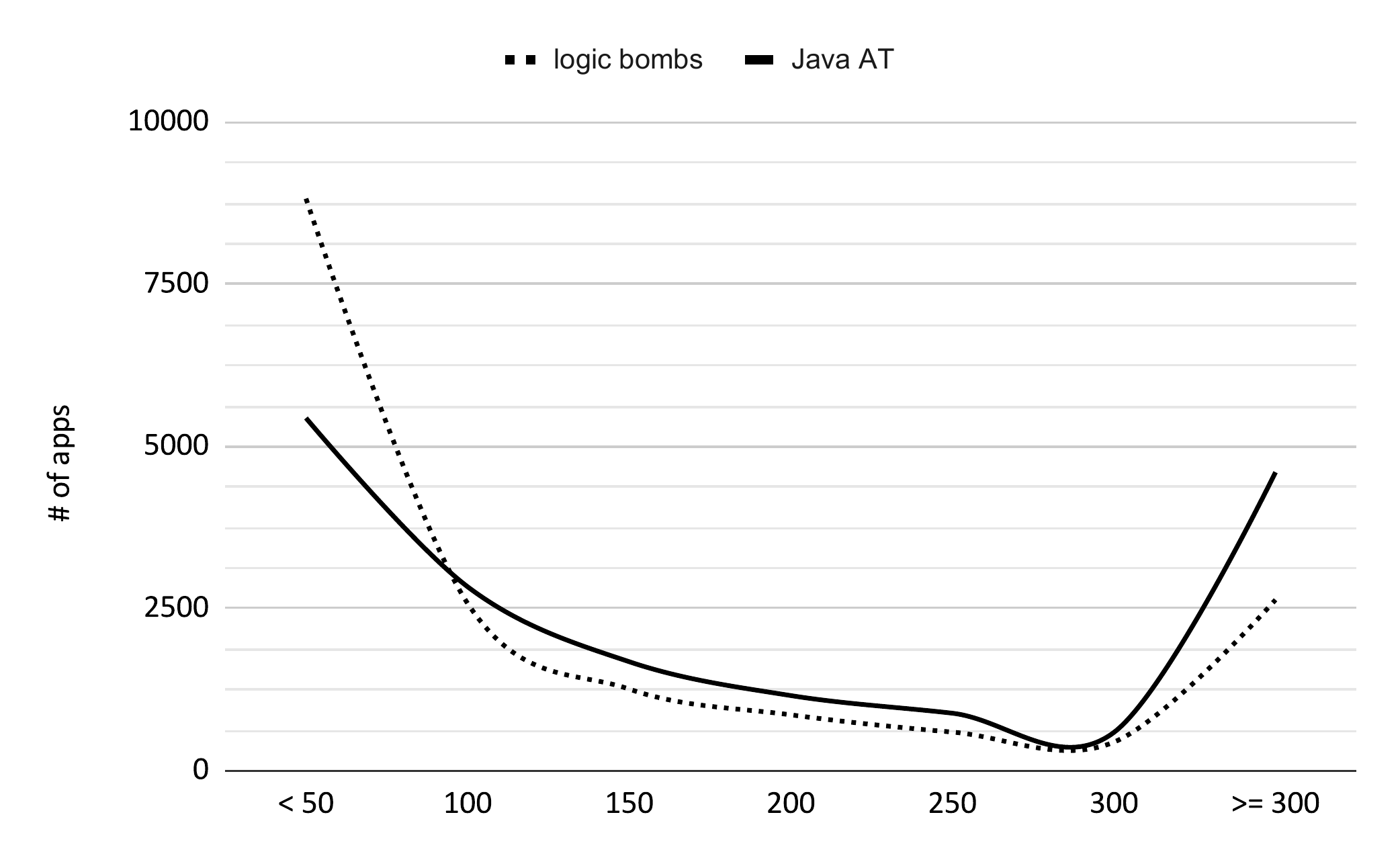}
  \caption{Java AT \& logic bomb}
  \label{fig:protection-values-recap-java-at}
\end{subfigure}%
\begin{subfigure}{.5\textwidth}
  \centering
  \includegraphics[width=\linewidth]{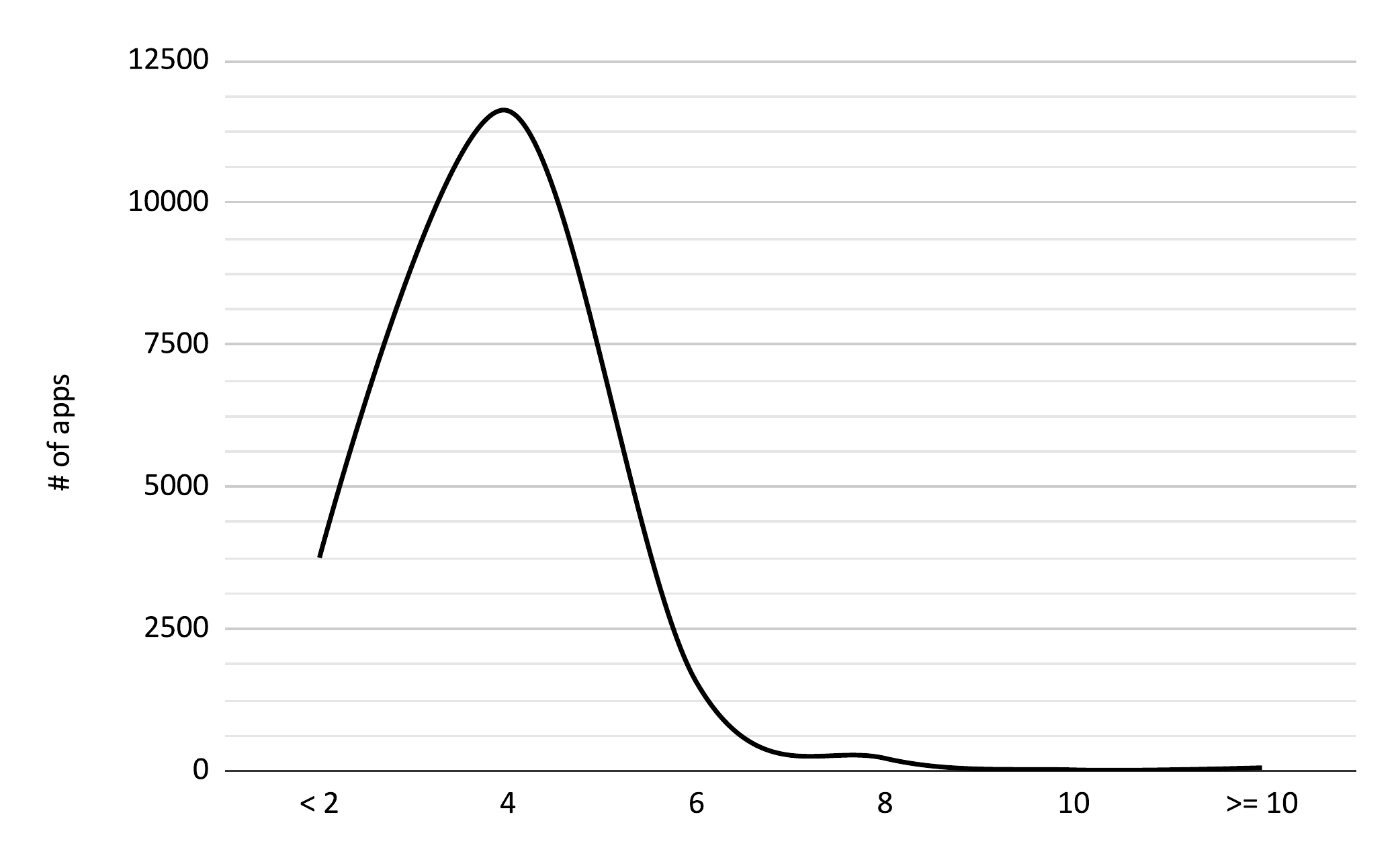}
  \caption{Nesting level}
  \label{fig:protection-values-recap-nesting}
\end{subfigure}
\caption{Distribution of Java AT, logic bombs and nesting level of the protected apps (30.000 apps).}
\label{fig:protection-values-recap}
\end{figure}

\begin{figure}
\centering
\begin{subfigure}{.5\textwidth}
  \centering
  \includegraphics[width=\linewidth]{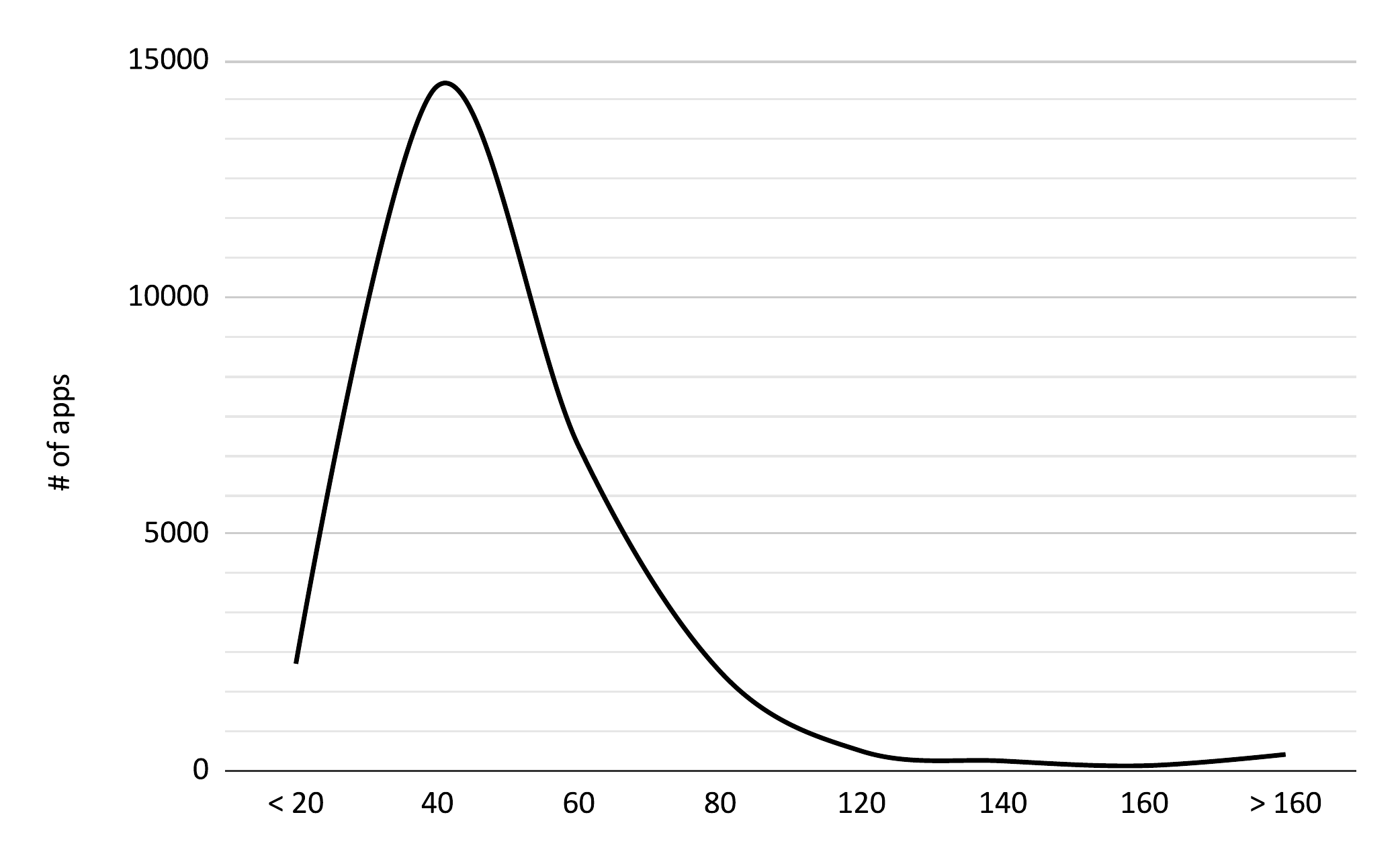}
  \caption{Processing time (sec)}
  \label{fig:overhead-recap-time}
\end{subfigure}%
\begin{subfigure}{.5\textwidth}
  \centering
  \includegraphics[width=\linewidth]{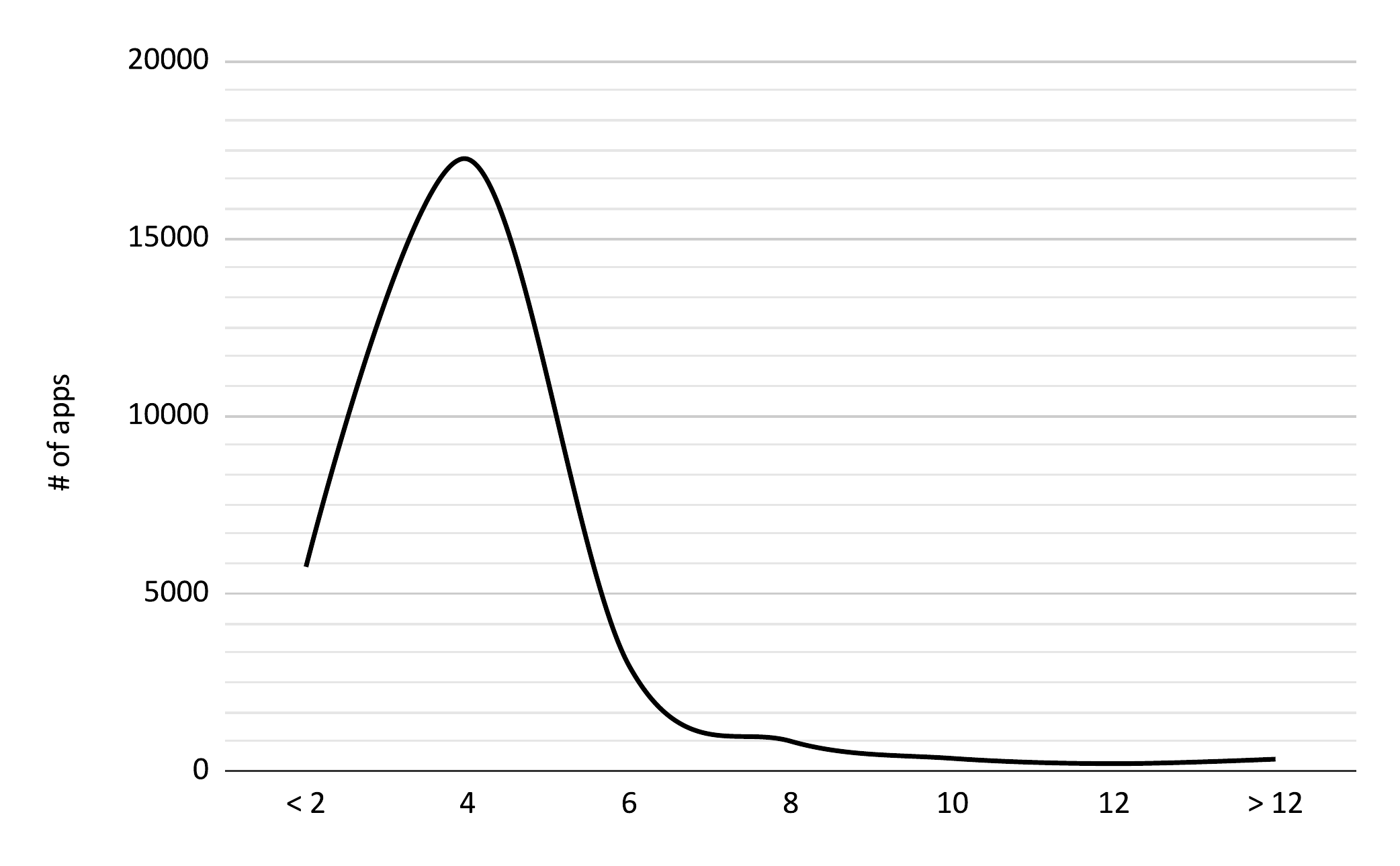}
  \caption{Size overhead (MB)}
  \label{fig:overhead-recap-space}
\end{subfigure}
\caption{Distribution of processing time and size overhead throw of the protected apps (30.000 apps).}
\label{fig:overhead-recap}
\end{figure}

\paragraph{\textbf{Protection Level.}} 
Figure \ref{fig:protection-values-recap} presents the distribution of the protection mechanisms (i.e., Java AT, logic bombs, and the nesting level) injected by \toolname{} over the set of the protected apps.
The results show that \toolname{} successfully includes a fair amount of logic bombs (i.e., 173 on average - distributed among the six possible types) and Java AT checks (i.e., 355 on average). 
Furthermore, the average nesting level of the logic bombs is greater than two (i.e., 2.3), as shown in Figure \ref{fig:protection-values-recap-nesting}.

\paragraph{\textbf{Processing Time.}} 
\toolname{} took 1.228.009 seconds (14 days and 5 hours) for protecting all the apps, with an average time of 48 seconds per app (cf. Figure \ref{fig:overhead-recap-time}), thereby suggesting that the approach is viable. 
The building time is mostly under 2 minutes for most of the analyzed apps (i.e., 26.103 apps), even for those containing many qualified conditions. 
During the experimental evaluation, only 229 reached the timeout. 

\paragraph{\textbf{Space Overhead.}}
The total space overhead is due to two factors: a fixed amount (calculated as less than 1MB) introduced by the transformation and packaging processes of \toolname{} and a dynamic one related to the number of added detection nodes. 
As expected, the number of introduced Java AT controls and logic bombs is directly proportional to the space overhead since each security control contributes to increase the payload size of each method. 
Furthermore, since native functions are compiled into a shared object for each architecture and embedded into the app, the space overhead is also affected by the number of native controls. 

Still, Figure \ref{fig:overhead-recap-space} shows that the vast majority of apps (i.e., 25.938 out of 27.656) have a size overhead less than 6MB, with 4.3 MB on average. All in all, it is worth noticing that only 41 apps violate the size limit of 100MB, imposed by several app stores, such as Google Play \cite{limit-size-android-apk}. For these apps, the average size overhead is 7.5MB, which means that the original apps already have a size near the limit (i.e., the average initial size is 96.7MB).

\subsection{Parameters Tuning}
\label{sec:parameter_tuning}

As \toolname{} uses three input parameters (i.e., $P_{Java\_AT}$, $P_{native\_key}$, and $P_{native\_AT}$) to determine the probability to include the Java AT checks and the six types of logic bombs, 
we conducted tuning tests to detect the on-average best combination of parameters to ensure a reasonable trade-off between the space and time overhead and the protection level. 

To do so, we computed fifteen different combinations of $P_{Java\_AT}, P_{native\_key}$, $P_{native\_AT}$. We used a testbed of 1.000 randomly-selected Android apps to evaluate i) the average number of injected Java AT and logic bombs, ii) the average nesting level, iii) the time required for the computation, and iv) the size overhead of the protected apk.
The results of the assessment, reported in Table \ref{table:params_compare}, lead to some considerations. For the evaluations, we set the $PN$ as $ApplicationID$. 

\begin{table}[]
\begin{tabular}{|c|c|c||c|c|c|c|c|}
\hline
\multicolumn{1}{|c|}{\textbf{\cellcolor[HTML]{CCFFFF}\begin{tabular}[c]{@{}c@{}}\% Java\\ AT\end{tabular}}} &
  \multicolumn{1}{c|}{\textbf{\cellcolor[HTML]{CCFFFF}\begin{tabular}[c]{@{}c@{}}\% Native\\ key\end{tabular}}} &
  \multicolumn{1}{c||}{\textbf{\cellcolor[HTML]{CCFFFF}\begin{tabular}[c]{@{}c@{}}\% Native\\ AT\end{tabular}}} &
  \multicolumn{1}{c|}{\textbf{\cellcolor[HTML]{FFFC9E}\begin{tabular}[c]{@{}c@{}}Avg. Java\\ AT\end{tabular}}} &
  \multicolumn{1}{c|}{\textbf{\cellcolor[HTML]{FFFC9E}Avg. LB}} &
  \multicolumn{1}{c|}{\textbf{\cellcolor[HTML]{FFFC9E}\begin{tabular}[c]{@{}c@{}}Avg. \\Nesting\end{tabular}}} &
  \multicolumn{1}{c|}{\textbf{\cellcolor[HTML]{FFFC9E}\begin{tabular}[c]{@{}c@{}}Avg. Protection \\ time (s)\end{tabular}}}&
  \multicolumn{1}{c|}{\textbf{\cellcolor[HTML]{FFFC9E}\begin{tabular}[c]{@{}c@{}}Avg. Size \\ overhead (MB)\end{tabular}}} \\ \hline
\multirow{4}{*}{5} & 30 & 30 & 87 & 95 & 2.4 & 22 & 2.44 \\\cline{2-8}
     & 80 & 20 & 86 & 92 & 2.2 & 23 & 2.73 \\\cline{2-8}
     & 60 & 40 & 86 & 94 & 2.1 & 23 & 2.66 \\\cline{2-8}
     & 40 & 60 & 87 & 95 & 2.1 & 22 & 2.67 \\\cline{2-8}
     & 20 & 80 & 87 & 95 & 2.1 & 23 & 2.66 \\\hline
\multirow{4}{*}{10} & 30 & 30 & 174 & 92 & 2.2 & 23 & 2.56 \\\cline{2-8}
     & 80 & 20 & 172 & 88 & 2.1 & 24 & 2.72 \\\cline{2-8}
     & 60 & 40 & 170 & 89 & 2.1 & 23 & 2.68\\\cline{2-8}
     & 40 & 60 & 174 & 91 & 2.1 & 23 & 2.67 \\\cline{2-8}
     & 20 & 80 & 171 & 90 & 2.1 & 23 & 2.66 \\\hline
\multirow{4}{*}{20} & 30 & 30 & 348 & 85 & 2.1 & 25 & 2.58 \\\cline{2-8}
     & 80 & 20 & 344 & 82 & 2.1 & 26 & 2.77 \\\cline{2-8}
     & 60 & 40 & 345 & 83 & 2.1 & 25 & 2.69\\\cline{2-8}
     & 40 & 60 & 349 & 84 & 2.1 & 25 & 2.68 \\\cline{2-8}
     & 20 & 80 & 345 & 84 & 2.0 & 24 & 2.68 \\\hline
\end{tabular}
\caption{Results of the input parameters tuning experiment (1.000 apps).}
\label{table:params_compare}
\end{table}

In all cases, the average nesting level of bombs is greater than 2, while the average number of included \toolname{} detection nodes (i.e., the sum of Java AT and the six types of logic bombs) ranges from 178 to 440. Such results ensure that all the apps are equipped with complex and hard-to-remove bombs despite the different input parameters. 
Regarding the space overhead, the results underline that the variation of the input parameters does not significantly affect the size of the protected apps: most apps have a size overhead of less than 6MB, with an average value around 3MB.

Finally, the experiments underline that most of the overhead in the processing time is due to the percentage of Java anti-tampering checks ($P_{Java\_AT}$). 
On the contrary, the percentage of native key ($P_{native\_key}$) and native AT ($P_{native\_AT}$) do not significantly affect the compilation time. 
To this aim, a reasonable configuration should use a low percentage value of Java AT (e.g., 10\%), but can set a higher percentage of native AT and native key bombs (e.g., between 40\% and 60\%).

\subsection{Runtime Performance}\label{subsec:effectiveness}

The last set of experiments aims at evaluating the effectiveness and usability of the \schema{} protection scheme in real-world Android apps.

To do so, we randomly-selected 200 apps protected with \toolname{}, and we analyzed them at runtime to check if the introduced protection does not affect the standard behavior of the app. The selected apps have an average of 95 Java AT and 50 logic bombs to ensure relevant results. 
Furthermore, we also tested the corresponding original apps to evaluate the runtime overhead introduced by the \schema{} protection.

For the testing phase, we used an emulated Android 8.0 phone equipped with a dual-core processor, 2GB of RAM, and the latest version of Google Play Services. We executed each app - original and protected one - for three minutes with the same high-level inputs (i.e., inputs that trigger an active element of the activity/view), and we retrieved the percentages of CPU and memory usages of each app every two high-level inputs. 

At the end of the experiments, 191 protected apps (i.e., 87\%) ran correctly without exceptions. The remaining 29 (i.e., 13\%) either thrown a new exception and crashed (i.e., 23 protected apps) or resulted in being unusable due to the high overhead introduced (i.e., 6 protected apps).

\begin{figure}
\centering
\begin{subfigure}{.5\textwidth}
  \centering
  \includegraphics[width=\linewidth]{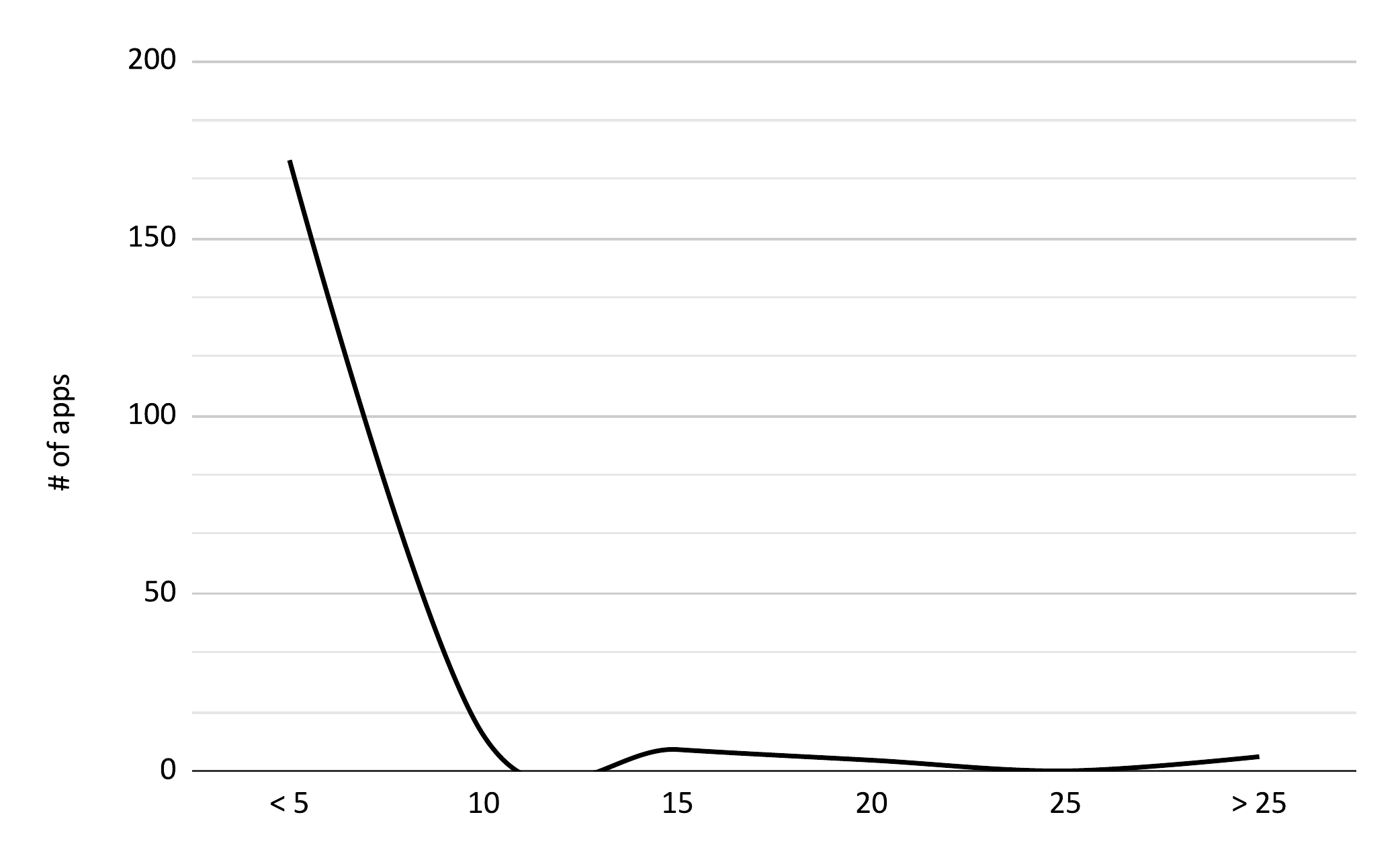}
  \caption{CPU usage overhead (\%)}
  \label{fig:cpu-overhead}
\end{subfigure}%
\begin{subfigure}{.5\textwidth}
  \centering
  \includegraphics[width=\linewidth]{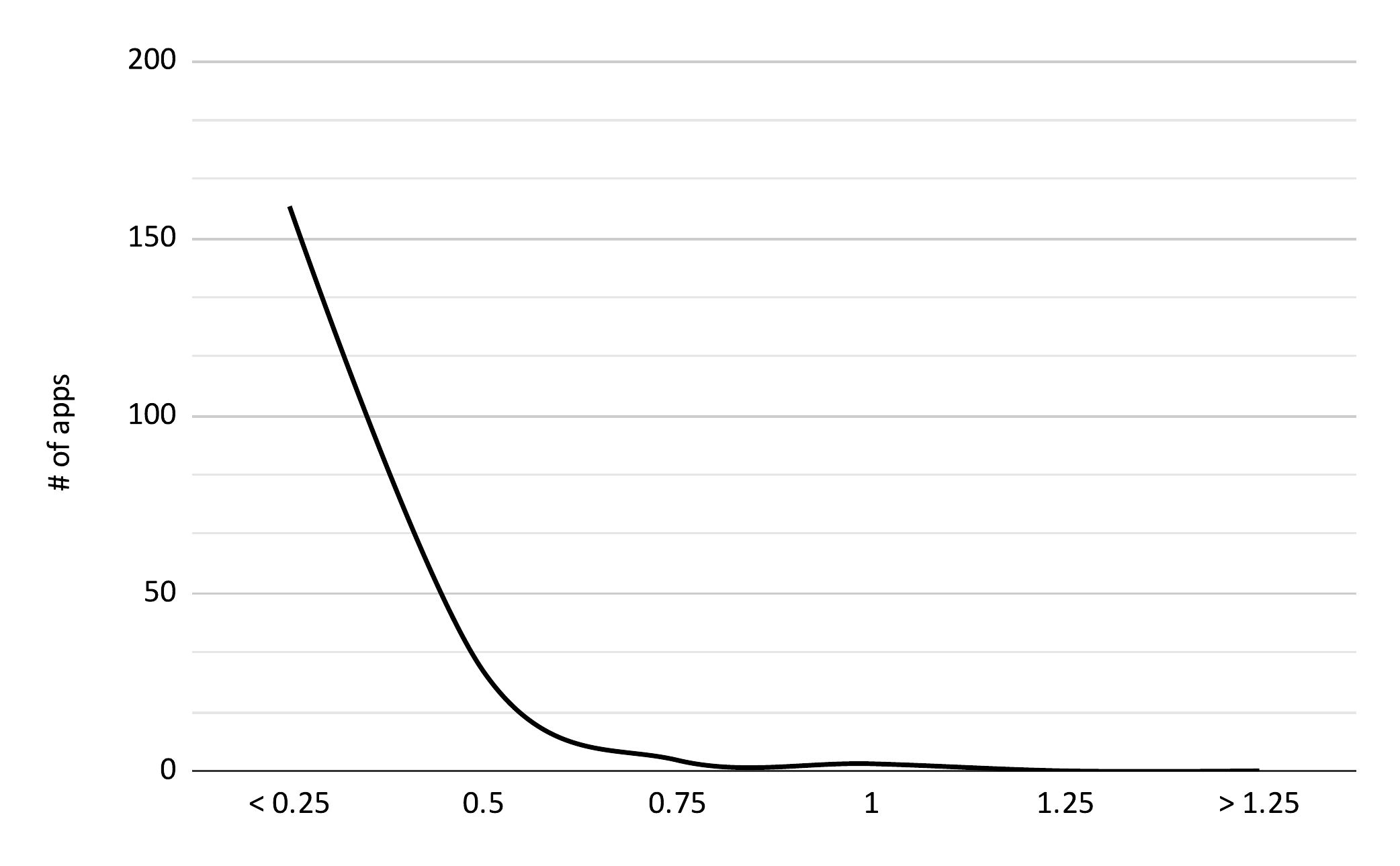}
  \caption{Memory usage overhead (\%)}
  \label{fig:memory-overhead}
\end{subfigure}
\caption{Distribution of the CPU and memory usage overhead during the runtime experiments (220 apps).}
\label{fig:overhead-runtime-recap}
\end{figure}

Figure \ref{fig:overhead-runtime-recap} reports the distribution of the CPU and memory overheads between the protected and the original app. 
The results show that the run-time overhead (i.e., CPU and memory usage) in the protected apps is negligible, with an average increase of 3\% in the CPU usage and 0.1\% in the memory usage and does not affect the user experience (i.e., no lags nor freezes).

Finally, it is worth noticing that the vast majority of the tested apps (i.e., 180 apps) have a CPU overhead less than 10\% and a memory overhead less than 0.5\%. 


\subsection{Prototype Limitations} \label{sec:limitation}

The experimental campaign led to the identification of some limitations of the \toolname{} tool that will be discussed below.
During the protection of 30.000 apps, 963 of them (i.e., 3.2\% of total apps) resulted in a crash due to the transformation in Jimple code. This is mainly related to the transformation of the switch-case statements. 
In particular, \toolname{} has to replace each switch-case statement with a list of if statements, and, thus, it has to update all links and dependencies to preserve the original execution flow. In some cases, \toolname{} fails to reconstruct the original dependencies causing a crash of the protection process.

Besides, \toolname{} inherits all the limitations of third-party libraries (e.g., the Soot framework) for both the protection phase and the runtime execution. In particular, during the transformation of the Java bytecode, Soot automatically applies some optimization techniques. Thus, the resulting bytecode, even the one outside $PN$, always differs from the original one. Unfortunately, such modifications in the case of third-party libraries (e.g., AdMob\footnote{\url{https://developers.google.com/admob/android/quick-start}}) may cause runtime errors that result in a user-experience glitch (e.g., a not-working advertisement banner).

Furthermore, \toolname{} leverages the \texttt{InMemoryDexClassLoader} class (which is introduced in API 26) to load and invoke the decrypted code at runtime. Due to this, the protected app needs to support API 26 as minimum API level or exceptions can occur\footnote{For instance, \url{https://stackoverflow.com/questions/38200282/android-os-fileuriexposedexception-file-storage-emulated-0-test-txt-exposed}} during its execution.

Finally, it is worth noticing that the protection depends on the portion of code that could be processed (i.e., the code contained in \texttt{PN}). Indeed, the selection of a package name with few classes and methods could result in an insufficient protection. For instance, some protected apps (6.115 out of the 27.656 apps) have not been equipped with any anti-tampering protection mechanism because the package name $ApplicationID$ contains few lines of Java bytecode (e.g., the case of hybrid apps). For this reason, the tool does not find any candidate qualified condition suitable to embed logic bombs.
However, it is worth noticing that such a problem can be mitigated by using a more permissive package name to cover, for instance, the entire app (i.e., setting \texttt{PN} to \emph{none}) and not only the $ApplicationID$ package.

\section{Conclusion and Future Works} \label{sec:conclusion}

In this work, we proposed \schema{}, an anti-repackaging protection scheme that overcomes the main limitations of current state-of-the-art protection schemes by introducing multiple protection patterns (i.e., Java anti-tampering checks and six different types of logic bombs) that relies on native technologies.

Furthermore, we implemented \schema{} in a tool for protecting Android apps, called \toolname{}, that is publicly available on GitHub\footnote{\url{https://github.com/Mobile-IoT-Security-Lab/ARMANDroid}} and on Docker Hub\footnote{\url{https://hub.docker.com/r/totor13/armand}}.
The evaluation of 30.000 Android apps demonstrated the applicability and the efficacy of the tool and the proposed protection scheme.
\toolname{} enables developers to protect their apps against repackaging attacks and can be easily included in existing application development processes. Furthermore, \toolname{} support the application, after the protection phase, of different obfuscation techniques (e.g., Obfuscapk \cite{aonzo2020obfuscapk}, ProGuard \cite{proguard}) to make static analysis harder for an attacker.

As a future extension of this work, we plan to \emph{i)} define a heuristic to evaluate and select the best QCs to be transformed into detection nodes, \emph{ii)} extend the multi-pattern protections (e.g., by adding mixed client-side and server-side controls), and \emph{iii)} evaluate the applicability to the iOS environment.

\bibliographystyle{model1-num-names}
\bibliography{main}

\begin{thebibliography}{29}
\expandafter\ifx\csname natexlab\endcsname\relax\def\natexlab#1{#1}\fi
\providecommand{\url}[1]{\texttt{#1}}
\providecommand{\href}[2]{#2}
\providecommand{\path}[1]{#1}
\providecommand{\DOIprefix}{doi:}
\providecommand{\ArXivprefix}{arXiv:}
\providecommand{\URLprefix}{URL: }
\providecommand{\Pubmedprefix}{pmid:}
\providecommand{\doi}[1]{\href{http://dx.doi.org/#1}{\path{#1}}}
\providecommand{\Pubmed}[1]{\href{pmid:#1}{\path{#1}}}
\providecommand{\bibinfo}[2]{#2}
\ifx\xfnm\relax \def\xfnm[#1]{\unskip,\space#1}\fi
\bibitem[{Desnos and Gueguen(2011)}]{2011-android-reversing}
\bibinfo{author}{A.~Desnos}, \bibinfo{author}{G.~Gueguen},
\newblock \bibinfo{title}{Android: From reversing to decompilation},
\newblock \bibinfo{journal}{Proc. of Black Hat Abu Dhabi}
  (\bibinfo{year}{2011}).
\bibitem[{Google(2020)}]{google-play-store}
\bibinfo{author}{Google}, \bibinfo{title}{Google play store},
  \bibinfo{year}{2020}. \URLprefix \url{https://play.google.com/store},
  \bibinfo{note}{accessed online: \today}.
\bibitem[{SAMSUNG ELECTRONICS~CO.(2020)}]{samsung-store}
\bibinfo{author}{L.~SAMSUNG ELECTRONICS~CO.}, \bibinfo{title}{Samsung store},
  \bibinfo{year}{2020}. \URLprefix
  \url{https://www.samsung.com/it/apps/galaxy-store/}, \bibinfo{note}{accessed
  online: \today}.
\bibitem[{Aonzo et~al.(2018)Aonzo, Merlo, Tavella, and
  Fratantonio}]{phishing-ccs18}
\bibinfo{author}{S.~Aonzo}, \bibinfo{author}{A.~Merlo},
  \bibinfo{author}{G.~Tavella}, \bibinfo{author}{Y.~Fratantonio},
\newblock \bibinfo{title}{Phishing attacks on modern android},
\newblock in: \bibinfo{booktitle}{Proceedings of the 2018 ACM SIGSAC Conference
  on Computer and Communications Security}, CCS '18,
  \bibinfo{publisher}{Association for Computing Machinery},
  \bibinfo{address}{New York, NY, USA}, \bibinfo{year}{2018}, p.
  \bibinfo{pages}{1788–1801}. \URLprefix
  \url{https://doi.org/10.1145/3243734.3243778}.
  \DOIprefix\doi{10.1145/3243734.3243778}.
\bibitem[{{Protsenko} et~al.(2015){Protsenko}, {Kreuter}, and
  {Müller}}]{7299906}
\bibinfo{author}{M.~{Protsenko}}, \bibinfo{author}{S.~{Kreuter}},
  \bibinfo{author}{T.~{Müller}},
\newblock \bibinfo{title}{Dynamic self-protection and tamperproofing for
  android apps using native code},
\newblock in: \bibinfo{booktitle}{2015 10th International Conference on
  Availability, Reliability and Security}, \bibinfo{year}{2015}, pp.
  \bibinfo{pages}{129--138}.
\bibitem[{{Luo} et~al.(2016){Luo}, {Fu}, {Wu}, {Zhu}, and {Liu}}]{7579771}
\bibinfo{author}{L.~{Luo}}, \bibinfo{author}{Y.~{Fu}},
  \bibinfo{author}{D.~{Wu}}, \bibinfo{author}{S.~{Zhu}},
  \bibinfo{author}{P.~{Liu}},
\newblock \bibinfo{title}{Repackage-proofing android apps},
\newblock in: \bibinfo{booktitle}{2016 46th Annual IEEE/IFIP International
  Conference on Dependable Systems and Networks (DSN)}, \bibinfo{year}{2016},
  pp. \bibinfo{pages}{550--561}.
\bibitem[{{Song} et~al.(2017){Song}, {Tang}, {Li}, {Gong}, {Chen}, {Fang}, and
  {Wang}}]{8368344}
\bibinfo{author}{L.~{Song}}, \bibinfo{author}{Z.~{Tang}},
  \bibinfo{author}{Z.~{Li}}, \bibinfo{author}{X.~{Gong}},
  \bibinfo{author}{X.~{Chen}}, \bibinfo{author}{D.~{Fang}},
  \bibinfo{author}{Z.~{Wang}},
\newblock \bibinfo{title}{Appis: Protect android apps against runtime
  repackaging attacks},
\newblock in: \bibinfo{booktitle}{2017 IEEE 23rd International Conference on
  Parallel and Distributed Systems (ICPADS)}, \bibinfo{year}{2017}, pp.
  \bibinfo{pages}{25--32}.
\bibitem[{{Chen} et~al.(2018){Chen}, {Zhang}, and {Liu}}]{8186215}
\bibinfo{author}{K.~{Chen}}, \bibinfo{author}{Y.~{Zhang}},
  \bibinfo{author}{P.~{Liu}},
\newblock \bibinfo{title}{Leveraging information asymmetry to transform android
  apps into self-defending code against repackaging attacks},
\newblock \bibinfo{journal}{IEEE Transactions on Mobile Computing}
  \bibinfo{volume}{17} (\bibinfo{year}{2018}) \bibinfo{pages}{1879--1893}.
\bibitem[{Zeng et~al.(2018)Zeng, Luo, Qian, Du, and Li}]{10.1145/3168820}
\bibinfo{author}{Q.~Zeng}, \bibinfo{author}{L.~Luo}, \bibinfo{author}{Z.~Qian},
  \bibinfo{author}{X.~Du}, \bibinfo{author}{Z.~Li},
\newblock \bibinfo{title}{Resilient decentralized android application
  repackaging detection using logic bombs},
\newblock in: \bibinfo{booktitle}{Proceedings of the 2018 International
  Symposium on Code Generation and Optimization}, CGO 2018,
  \bibinfo{publisher}{Association for Computing Machinery},
  \bibinfo{address}{New York, NY, USA}, \bibinfo{year}{2018}, p.
  \bibinfo{pages}{50–61}. \URLprefix \url{https://doi.org/10.1145/3168820}.
  \DOIprefix\doi{10.1145/3168820}.
\bibitem[{Tanner et~al.(2020)Tanner, Vogels, and
  Wattenhofer}]{10.1007/978-3-030-45371-8_12}
\bibinfo{author}{S.~Tanner}, \bibinfo{author}{I.~Vogels},
  \bibinfo{author}{R.~Wattenhofer},
\newblock \bibinfo{title}{Protecting android apps from repackaging using native
  code},
\newblock in: \bibinfo{editor}{A.~Benzekri}, \bibinfo{editor}{M.~Barbeau},
  \bibinfo{editor}{G.~Gong}, \bibinfo{editor}{R.~Laborde},
  \bibinfo{editor}{J.~Garcia-Alfaro} (Eds.), \bibinfo{booktitle}{Foundations
  and Practice of Security}, \bibinfo{publisher}{Springer International
  Publishing}, \bibinfo{address}{Cham}, \bibinfo{year}{2020}, pp.
  \bibinfo{pages}{189--204}.
\bibitem[{Merlo et~al.(2020)Merlo, Ruggia, Sciolla, and
  Verderame}]{merlo2020you}
\bibinfo{author}{A.~Merlo}, \bibinfo{author}{A.~Ruggia},
  \bibinfo{author}{L.~Sciolla}, \bibinfo{author}{L.~Verderame},
\newblock \bibinfo{title}{You shall not repackage! a journey into the world of
  anti-repackaging on android},
\newblock \bibinfo{journal}{Computers \& Security, to appear}
  (\bibinfo{year}{2020}).
\bibitem[{Micro(2012)}]{google-bouncer}
\bibinfo{author}{T.~Micro}, \bibinfo{title}{A look at google bouncer},
  \bibinfo{howpublished}{\url{https://www.trendmicro.com/en_us/research/12/g/a-look-at-google-bouncer.html}},
  \bibinfo{year}{2012}. \bibinfo{note}{Accessed online: \today}.
\bibitem[{Google(2012)}]{google-play-protect}
\bibinfo{author}{Google}, \bibinfo{title}{Google play protect},
  \bibinfo{howpublished}{\url{https://support.google.com/accounts/answer/2812853}},
  \bibinfo{year}{2012}. \bibinfo{note}{Accessed online: \today}.
\bibitem[{Gultnieks et~al.(2020)Gultnieks, Fuentes, Oberhauser, and
  Demirtaş}]{f-droid}
\bibinfo{author}{C.~Gultnieks}, \bibinfo{author}{A.~A. Fuentes},
  \bibinfo{author}{A.~Oberhauser}, \bibinfo{author}{A.~Demirtaş},
  \bibinfo{title}{F-droid}, \bibinfo{year}{2020}. \URLprefix
  \url{https://www.f-droid.org/}, \bibinfo{note}{accessed online: \today}.
\bibitem[{Google(2020{\natexlab{a}})}]{android-ndk}
\bibinfo{author}{Google}, \bibinfo{title}{Android ndk},
  \bibinfo{year}{2020}{\natexlab{a}}. \URLprefix
  \url{https://developer.android.com/ndk}, \bibinfo{note}{accessed online:
  \today}.
\bibitem[{Google(2020{\natexlab{b}})}]{android-publish-app}
\bibinfo{author}{Google}, \bibinfo{title}{Android: Publish your app},
  \bibinfo{howpublished}{\url{https://developer.android.com/studio/publish}},
  \bibinfo{year}{2020}{\natexlab{b}}. \bibinfo{note}{Accessed online: \today}.
\bibitem[{Sharif et~al.(2008)Sharif, Lanzi, Giffin, and
  Lee}]{sharif2008impeding}
\bibinfo{author}{M.~I. Sharif}, \bibinfo{author}{A.~Lanzi},
  \bibinfo{author}{J.~T. Giffin}, \bibinfo{author}{W.~Lee},
\newblock \bibinfo{title}{Impeding malware analysis using conditional code
  obfuscation},
\newblock in: \bibinfo{booktitle}{NDSS}, \bibinfo{year}{2008}.
\bibitem[{Reaves et~al.(2016)Reaves, Bowers, Gorski~III, Anise, Bobhate, Cho,
  Das, Hussain, Karachiwala, Scaife, Wright, Butler, Enck, and
  Traynor}]{10.1145/2996358}
\bibinfo{author}{B.~Reaves}, \bibinfo{author}{J.~Bowers},
  \bibinfo{author}{S.~A. Gorski~III}, \bibinfo{author}{O.~Anise},
  \bibinfo{author}{R.~Bobhate}, \bibinfo{author}{R.~Cho},
  \bibinfo{author}{H.~Das}, \bibinfo{author}{S.~Hussain},
  \bibinfo{author}{H.~Karachiwala}, \bibinfo{author}{N.~Scaife},
  \bibinfo{author}{B.~Wright}, \bibinfo{author}{K.~Butler},
  \bibinfo{author}{W.~Enck}, \bibinfo{author}{P.~Traynor},
\newblock \bibinfo{title}{*droid: Assessment and evaluation of android
  application analysis tools},
\newblock \bibinfo{journal}{ACM Comput. Surv.} \bibinfo{volume}{49}
  (\bibinfo{year}{2016}). \URLprefix \url{https://doi.org/10.1145/2996358}.
  \DOIprefix\doi{10.1145/2996358}.
\bibitem[{Li et~al.(2017)Li, Bissyandé, Papadakis, Rasthofer, Bartel, Octeau,
  Klein, and Traon}]{LI201767}
\bibinfo{author}{L.~Li}, \bibinfo{author}{T.~F. Bissyandé},
  \bibinfo{author}{M.~Papadakis}, \bibinfo{author}{S.~Rasthofer},
  \bibinfo{author}{A.~Bartel}, \bibinfo{author}{D.~Octeau},
  \bibinfo{author}{J.~Klein}, \bibinfo{author}{L.~Traon},
\newblock \bibinfo{title}{Static analysis of android apps: A systematic
  literature review},
\newblock \bibinfo{journal}{Information and Software Technology}
  \bibinfo{volume}{88} (\bibinfo{year}{2017}) \bibinfo{pages}{67 -- 95}.
  \URLprefix
  \url{http://www.sciencedirect.com/science/article/pii/S0950584917302987}.
  \DOIprefix\doi{https://doi.org/10.1016/j.infsof.2017.04.001}.
\bibitem[{Gajrani et~al.(2020)Gajrani, Laxmi, Tripathi, Gaur, Zemmari, Mosbah,
  and Conti}]{GAJRANI202073}
\bibinfo{author}{J.~Gajrani}, \bibinfo{author}{V.~Laxmi},
  \bibinfo{author}{M.~Tripathi}, \bibinfo{author}{M.~S. Gaur},
  \bibinfo{author}{A.~Zemmari}, \bibinfo{author}{M.~Mosbah},
  \bibinfo{author}{M.~Conti},
\newblock \bibinfo{title}{Chapter three - effectiveness of state-of-the-art
  dynamic analysis techniques in identifying diverse android malware and future
  enhancements},
\newblock volume \bibinfo{volume}{119} of \textit{\bibinfo{series}{Advances in
  Computers}}, \bibinfo{publisher}{Elsevier}, \bibinfo{year}{2020}, pp.
  \bibinfo{pages}{73 -- 120}. \URLprefix
  \url{http://www.sciencedirect.com/science/article/pii/S0065245820300413}.
  \DOIprefix\doi{https://doi.org/10.1016/bs.adcom.2020.03.002}.
\bibitem[{Viticchi\'{e} et~al.(2016)Viticchi\'{e}, Basile, Avancini, Ceccato,
  Abrath, and Coppens}]{10.1145/2995306.2995315}
\bibinfo{author}{A.~Viticchi\'{e}}, \bibinfo{author}{C.~Basile},
  \bibinfo{author}{A.~Avancini}, \bibinfo{author}{M.~Ceccato},
  \bibinfo{author}{B.~Abrath}, \bibinfo{author}{B.~Coppens},
\newblock \bibinfo{title}{Reactive attestation: Automatic detection and
  reaction to software tampering attacks},
\newblock in: \bibinfo{booktitle}{Proceedings of the 2016 ACM Workshop on
  Software PROtection}, SPRO '16, \bibinfo{publisher}{Association for Computing
  Machinery}, \bibinfo{address}{New York, NY, USA}, \bibinfo{year}{2016}, p.
  \bibinfo{pages}{73–84}. \URLprefix
  \url{https://doi.org/10.1145/2995306.2995315}.
  \DOIprefix\doi{10.1145/2995306.2995315}.
\bibitem[{Wang et~al.(2018)Wang, Liu, Gao, Jingjing, and
  Liu}]{10.1007/978-3-030-05234-8_16}
\bibinfo{author}{T.~Wang}, \bibinfo{author}{L.~Liu}, \bibinfo{author}{C.~Gao},
  \bibinfo{author}{H.~Jingjing}, \bibinfo{author}{J.~Liu},
  \bibinfo{title}{Towards Android Application Protection via Kernel Extension:
  ICA3PP 2018 International Workshops, Guangzhou, China, November 15-17, 2018,
  Proceedings}, \bibinfo{year}{2018}, pp. \bibinfo{pages}{131--137}.
  \DOIprefix\doi{10.1007/978-3-030-05234-8_16}.
\bibitem[{{Buhov} et~al.(2015){Buhov}, {Huber}, {Merzdovnik}, {Weippl}, and
  {Dimitrova}}]{7299933}
\bibinfo{author}{D.~{Buhov}}, \bibinfo{author}{M.~{Huber}},
  \bibinfo{author}{G.~{Merzdovnik}}, \bibinfo{author}{E.~{Weippl}},
  \bibinfo{author}{V.~{Dimitrova}},
\newblock \bibinfo{title}{Network security challenges in android applications},
\newblock in: \bibinfo{booktitle}{2015 10th International Conference on
  Availability, Reliability and Security}, \bibinfo{year}{2015}, pp.
  \bibinfo{pages}{327--332}. \DOIprefix\doi{10.1109/ARES.2015.59}.
\bibitem[{{Bhor} and {Karia}(2017)}]{8068748}
\bibinfo{author}{M.~{Bhor}}, \bibinfo{author}{D.~{Karia}},
\newblock \bibinfo{title}{Certificate pinning for android applications},
\newblock in: \bibinfo{booktitle}{2017 International Conference on Inventive
  Systems and Control (ICISC)}, \bibinfo{year}{2017}, pp.
  \bibinfo{pages}{1--4}. \DOIprefix\doi{10.1109/ICISC.2017.8068748}.
\bibitem[{Aonzo et~al.(2020)Aonzo, Georgiu, Verderame, and
  Merlo}]{aonzo2020obfuscapk}
\bibinfo{author}{S.~Aonzo}, \bibinfo{author}{G.~C. Georgiu},
  \bibinfo{author}{L.~Verderame}, \bibinfo{author}{A.~Merlo},
\newblock \bibinfo{title}{Obfuscapk: An open-source black-box obfuscation tool
  for android apps},
\newblock \bibinfo{journal}{SoftwareX} \bibinfo{volume}{11}
  (\bibinfo{year}{2020}) \bibinfo{pages}{100403}. \URLprefix
  \url{http://www.sciencedirect.com/science/article/pii/S2352711019302791}.
  \DOIprefix\doi{https://doi.org/10.1016/j.softx.2020.100403}.
\bibitem[{Group(2020)}]{soot}
\bibinfo{author}{S.~R. Group}, \bibinfo{title}{Soot - a java optimization
  framework}, \bibinfo{year}{2020}. \URLprefix
  \url{https://github.com/soot-oss/soot}, \bibinfo{note}{accessed online:
  \today}.
\bibitem[{Vallee-Rai and Hendren(1998)}]{vallee1998jimple}
\bibinfo{author}{R.~Vallee-Rai}, \bibinfo{author}{L.~J. Hendren},
\newblock \bibinfo{title}{Jimple: Simplifying java bytecode for analyses and
  transformations}  (\bibinfo{year}{1998}).
\bibitem[{Google(2020)}]{limit-size-android-apk}
\bibinfo{author}{Google}, \bibinfo{title}{Android developer},
  \bibinfo{year}{2020}. \URLprefix
  \url{https://support.google.com/googleplay/android-developer/answer/113469?hl=en},
  \bibinfo{note}{accessed online: \today}.
\bibitem[{Guardsquare(2020)}]{proguard}
\bibinfo{author}{Guardsquare}, \bibinfo{title}{Proguard}, \bibinfo{year}{2020}.
  \URLprefix
  \url{https://www.guardsquare.com/en/products/proguard/manual/introduction},
  \bibinfo{note}{accessed online: \today}.

\end{thebibliography}

\newpage

\bio{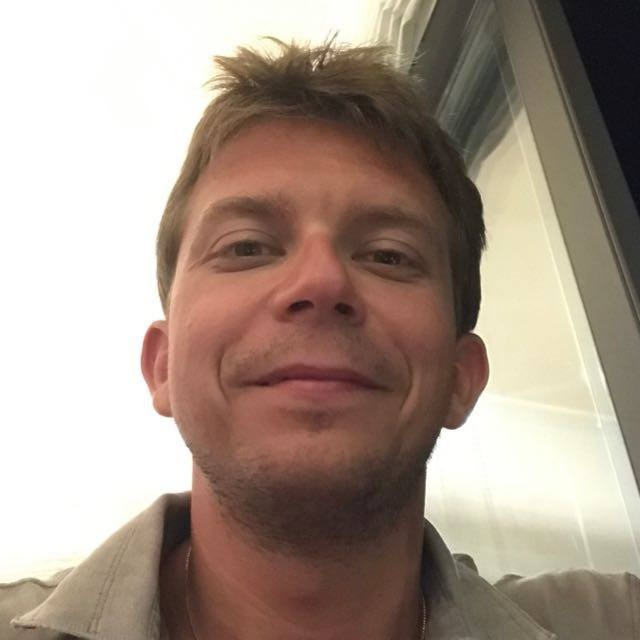}
\textbf{Alessio Merlo} is an Associate Professor in Computer Engineering in the Departiment of Informatics, Bioengineering, Robotics and System Engineering Department (DIBRIS) at the University of Genoa, and a member of the Computer Security Laboratory (CSEC Lab). His main research field is Mobile Security, with a specific interest on Android security, automated static and dynamic analysis of Android apps, mobile authentication and mobile malware. More information can be found at: \url{http://csec.it/people/alessio\_merlo/} 
\endbio

\vspace{0.2in}

\bio{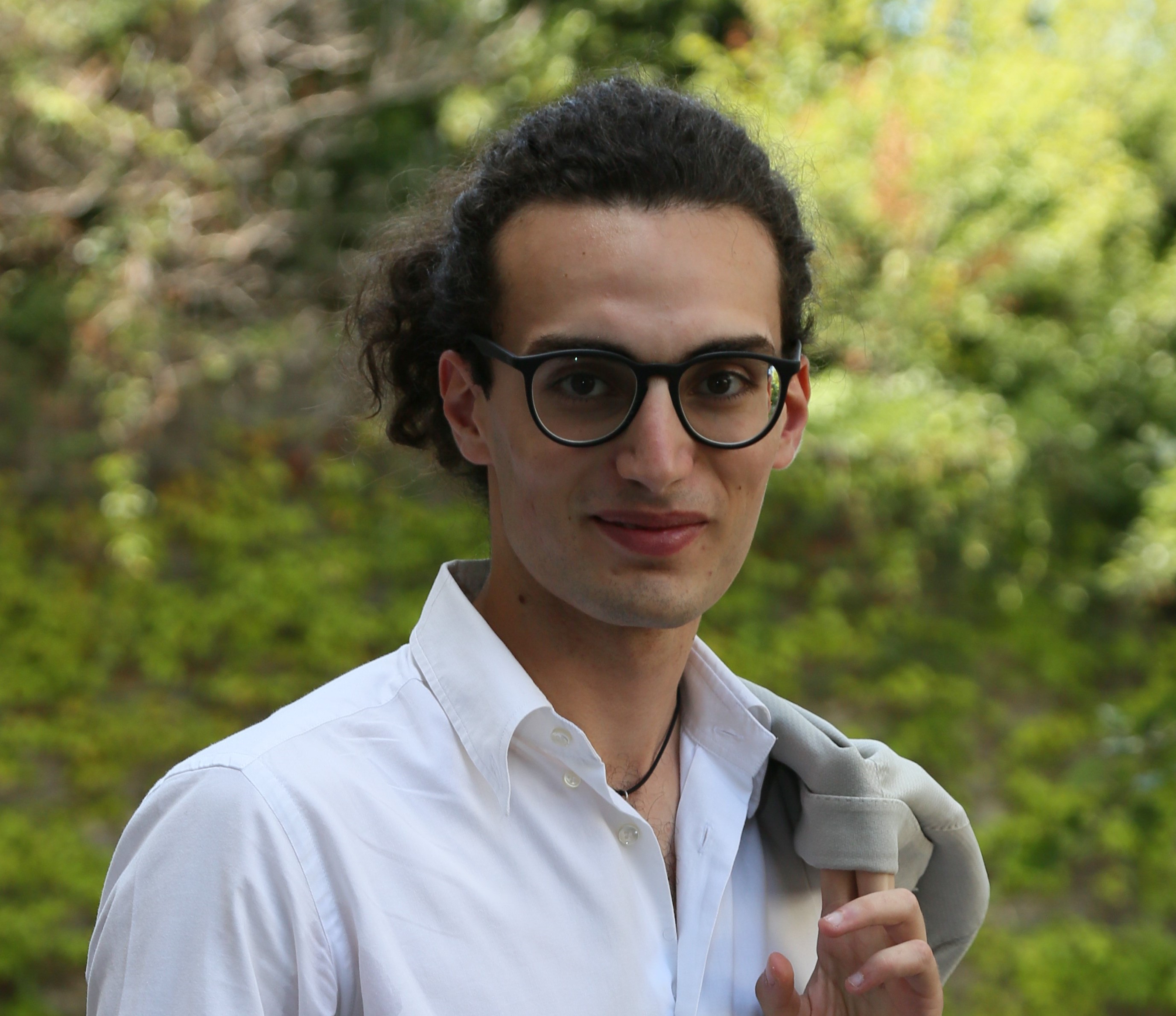}
\textbf{Antonio Ruggia} is a PhD student in Computer Engineering at the University of Genoa. He is interested in several security topics that include Mobile Security, with a specific interest in Android and data protection. He graduated in October 2020 at University of Genoa and participated in the 2019 CyberChallenge.it, an Italian practical competition for students in Cybersecurity. He regularly participates in international Capture-The-Flag (CTF) competitions. Since 2018, he has been working as a full-stack developer in a multinational corporation.
\endbio

\vspace{0.1in}

\bio{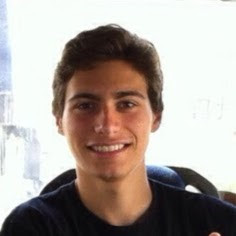}
\textbf{Luigi Sciolla} holds a Master degree in Computer Engineering at University of Genoa. During his studies he worked as a system administrator and full stack mobile developer. His area of interests are computer networks and Android security. He participated in the 2018 CyberChallenge.IT finals. Since 2018, he is also member of ZenHack, the capture-the-flag team at the University of Genoa. With ZenHack he competed in dozens of online and onsite competitions. 
\endbio

\vspace{0.25in}

\bio{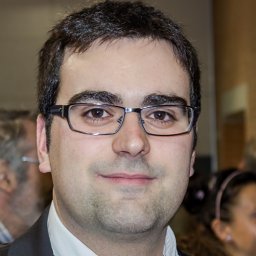}
\textbf{Luca Verderame} obtained a Ph.D. in Electronic, Information, Robotics, and Telecommunication Engineering at the University of Genoa (Italy) in 2016, where he worked on mobile security. He is currently working as a post-doc research fellow at the Computer Security Laboratory (CSEC Lab), and he is also the CEO and Co-founder of Talos, a cybersecurity startup and university spin-off. His research interests mainly cover information security applied, in particular, to mobile and IoT environments.
\endbio

\end{document}